\documentclass[12pt]{article}
\usepackage[english]{babel}
\usepackage{amssymb}
\usepackage{graphicx}
\usepackage{epsfig}

\def\a{\alpha}

\def\c{\gamma}
\def\b{\beta}
\def\d{\delta}

\def\+{\bigoplus}
\def\D{\Delta}

\def\({\left(}
\def\){\right)}
\def\[{\left[}
\def\]{\right]}
\def\l.{\left.}
\def\r.{\right.}

\def\be{\begin{equation}}
\def\ee{\end{equation}}
\def\bea{\begin{eqnarray}}
\def\eea{\end{eqnarray}}

\def\ber{\begin{array}}
\def\eer{\end{array}}

\def\scl{{\scriptstyle(}}
\def\scr{{\scriptstyle)}}
\def\Ddot{\dot{\D}_\Lambda}
\def\Del{\D_\Lambda}
\def\Dinv{\D^{-1}_\Lambda}
\def\sss{\scriptscriptstyle}

\def\Gammapis{{\widetilde\Gamma}}

\begin{document}

\begin{titlepage}
\rightline{GEF-TH-3-2003 }
\vskip 1in
\begin{center}
\def\thefootnote{\fnsymbol{footnote}}
{\large \bf The Renormalization of Non-Commutative Field Theories in the
Limit of Large Non-Commutativity} 
\vskip 0.3in
C. Becchi \footnote{E-Mail: becchi@ge.infn.it}, 
S. Giusto \footnote{E-Mail: giusto@ge.infn.it} 
and C. Imbimbo \footnote{E-Mail: imbimbo@ge.infn.it} 
\vskip .2in
{\em Dipartimento di Fisica, Universit\`a di Genova\\
and\\ Istituto Nazionale di Fisica Nucleare, Sezione di Genova\\
via Dodecaneso 33, I-16146, Genoa, Italy}
\end{center}
\vskip .2in
\begin{abstract}
We show that renormalized non-commutative scalar field theories do not
reduce to their planar sector in the limit of large non-commutativity.
This follows from the fact that the RG equation of the
Wilson-Polchinski type which describes the genus zero sector of
non-commutative field theories couples generic planar amplitudes with
non-planar amplitudes at exceptional values of the external
momenta. We prove that the renormalization problem can be consistently
restricted to this set of amplitudes. In the resulting renormalized
theory non-planar divergences are treated as UV divergences requiring
appropriate non-local counterterms. In 4 dimensions the model turns
out to have one more relevant (non-planar) coupling than its
commutative counterpart. This non-planar coupling is ``evanescent'':
although in the massive (but not in the massless) case its
contribution to planar amplitudes vanishes when the floating cut-off
equals the renormalization scale, this coupling is needed to make the
Wilsonian effective action UV finite at {\it all} values of the
floating cut-off.
\end{abstract}
\vfill
\end{titlepage}
\setcounter{footnote}{0}

\section{Introduction}

The aim of this paper is to address some issues which arise in the
renormalization of non-commutative quantum field theories in the limit
when the non-commutativity parameter $\theta$ is large. Feynman
diagrams of non-commutative theories, like those of matrix field
theories, have a double line representation and thus admit a
topological classification in terms of oriented Riemann surfaces with
holes to which external lines are attached.  Diagrams with spherical
topology are called {\it planar} when they have a single hole to which
all the external lines are attached --- in the matrix theories these
are also called single trace diagrams.  {\it Non-planar} spherical
diagrams have more than one hole and, in the matrix models, correspond
to multi-trace terms of the effective action.

The current understanding of the renormalization of non-commutative
theories is based on the observation that planar diagrams have
exactly the same divergences as in the commutative theory.
Divergences of non-planar graphs are instead 
automatically regulated, in the non-commutative theory, by an
effective UV cut-off $1/\theta \,p$, where $p$ is the momentum
entering a hole of the diagram. Since the
effective UV cut-off diverges when the momentum $p$ entering
a hole vanishes, non-planar diagrams diverge when evaluated
at {\it exceptional} values of the external momenta --- the famous 
IR/UV mixing effect. Therefore it has been conjectured
\cite{seiberg} that to remove all UV divergences of non-commutative 
amplitudes at {\it generic} values of the external momenta,
it is sufficient to introduce counterterms corresponding
to planar divergences only. In the following we will refer to this
as the ``planar renormalization'' scheme of non-commutative theories.
Explicit computations up to two loops have been performed that seem to
confirm this expectation~\cite{arefeva,micu}.

That planar renormalization should work is not ``a priori'' obvious  
and might in fact even appear to be surprising, since planar amplitudes 
have in general non-planar subgraphs: these subgraphs
necessarily appear at exceptional values of their external momenta and thus,
in planar renormalization, may lead to unsubtracted divergences. 
In this paper we will be able to explain when and in which ---
limited --- sense planar renormalization ``works''.
 
The technical tool that we use to investigate the renormalization of
non-commutative theories is the Wilson-Polchinski renormalization
group equation that we derived in \cite{bgi}.  This equation, which
applies both to the large $N$ limit of matrix field theories and to the
large $\theta$ limit of non-commutative theories, describes the RG evolution
of amplitudes with ${\it spherical}$ topology. It makes manifest the
impossibility of limiting the renormalization problem to planar
diagrams: the RG evolution of a generic planar amplitude involves
necessarily non-planar spherical amplitudes. However, the non-planar
amplitudes that are coupled by the RG flow to the planar ones are not
generic --- they are restricted to momenta configurations for which
the total momenta $p$ entering each hole of the non-planar amplitude
vanish. We will refer to the amplitudes restricted to such
exceptional momenta as the {\it Partially Integrated Spherical} (PIS)
amplitudes: in configuration space they are Green functions
integrated over the centers of mass of all the points attached to the
same hole. PIS amplitudes include both planar amplitudes evaluated
at generic momenta and non-planar amplitude taken at exceptional
momenta. We see that the RG approach to renormalization of non-commutative theories
naturally leads to consider a special class of non-local observables, 
corresponding to PIS amplitudes. It should be kept in mind that
renormalized non-planar PIS amplitudes cannot be considered as
limits --- for $p\to 0$ ---  of generic non-planar spherical amplitudes.
However, since the RG equation closes over PIS amplitudes 
the renormalization problem for this set of amplitudes  
is well formulated in the Wilson-Polchinski
framework \cite{polchinski}. This is the problem that we will solve
in this paper by showing that the theory
of PIS amplitudes of non-commutative field theory
is renormalizable\footnote{The PIS sector of matrix field theory also
defines a consistent renormalization problem. However for matrix field
theory one can as well consider the renormalization of generic
spherical amplitudes.}: both in the sense that renormalized amplitudes
are finite when the the UV cut-off is removed and in the Wilsonian
sense that the Wilson-Polchinski effective action
is independent of the UV scale for any value of the floating cut-off $\Lambda$.

In conclusion, PIS theory is the renormalizable theory which describes
the $\theta=\infty$ limit of non-commutative field theory. 
We believe, although we do not address this issue in this paper,
that it also encodes the whole UV non-trivial content of the 
non-commutative theory at finite $\theta$: in other words we think
that, once PIS sub-divergences have been subtracted, the only
divergences left are to be treated as IR ones, as suggested in \cite{seiberg}.

The difference between planar
renormalization and our renormalization scheme is illuminated by a
factorization property of the $\theta\to\infty$ limit of
non-commutative theory that is the direct analogue of large $N$
factorization of matrix models.  Factorization follows from the fact
that the RG equation for $\theta\to\infty$ --- unlike the ordinary
commutative RG equation --- is of {\it first order} in source
derivatives. It is because of factorization that, when the Polchinski
floating cut-off $\Lambda$ equals the mass renormalization scale, 
the renormalized planar amplitudes depend on one less
marginal coupling than generic non-planar PIS amplitudes.  In this
sense one can say that the non-planar coupling is {\it evanescent}. It
turns out that in the {\it massive} theory one can neglect the non-planar
coupling if one only looks at planar amplitudes at
$\Lambda=\Lambda_R$, where $\Lambda_R$ is
the renormalization scale: in other words, 
{\it when the floating IR cut-off $\Lambda$ equals $\Lambda_R$} the planar
part of our Wilsonian effective action coincides with the effective
action that one would obtain from planar renormalization.
However, as soon as $\Lambda$ differs from $\Lambda_R$ the planar
effective action obtained from PIS theory and the one which neglects
non-planar divergences begin to differ from one another: the ``naive''
planar effective action becomes dependent on the UV scale $\Lambda_0$ while 
the effective action coming from PIS theory does
not\footnote{The planar part of the Wilsonian action obtained by 
planar renormalization  depends on the UV scale $\Lambda_0$ when the 
external momenta are  {\it generic}, as we will show in Section 2
by a specific computation. This disagrees with the opposite claim
made in \cite{griguolo}.}.  
The fact that the ``naive'' planar effective action 
depends on $\Lambda_0$ when $\Lambda\not=\Lambda_R$
might, at first sight, appear surprising since the Wilson-Polchinski 
RG equation is essentially independent of the UV scale $\Lambda_0$: 
the reason why this happens is that the RG equation does not close on
planar amplitudes and the $\Lambda$ derivative of a planar amplitude
involves non-planar diagrams evaluated at exceptional momenta which
are, in planar renormalization, divergent.
Hence the non-trivial dependence on $\Lambda_0$
of the ``naive'' planar effective action is the  ``shadow'' 
at the planar level of the IR/UV difficulty that afflicts non-planar 
amplitudes computed in planar renormalization.

Our renormalization framework can also be applied to the 
massless theory: this theory is particularly interesting since its
UV and IR divergences conspire
to produce an anomalous dependence of the planar amplitudes on the
non-planar coupling at $\Lambda=0$. Had one neglected
non-planar counterterms in the massless case,   
one would have obtained an effective planar action UV divergent 
for {\it any} value of $\Lambda$, including when $\Lambda\to 0$.

The reason why non-planar counterterms can be introduced in PIS theory 
is that non-planar PIS amplitudes depend on the non-commutative parameter $\theta$ via an overall Moyal phase. More precisely, let 
the Moyal phase of a planar diagram be  
\be
{\rm e}^{-i\, \Phi_n(p_1,\ldots, p_n)} \equiv {\rm e}^{-i\sum_{i<j} p_i
\wedge p_j}
\label{moyalphase}
\ee
where $p_1\wedge p_2 \equiv {1\over 2}\theta_{\mu\nu}\,p_1^\mu\, p_2^\nu$
and $p_1,\ldots,p_n$ are the momenta associated to the $n$ external
lines of the graph\footnote{We will
assume $\theta_{\mu\nu}$ to be a non-degenerate anti-symmetric matrix
and we will consider the euclidean theory.}. 
In Appendix A it will be shown that
a PIS amplitude
with $h$ holes depend on the $\theta$ via an overall factor
which is the product of $h$ factors like (\ref{moyalphase}) ---
one for each hole. This should be contrasted with the complicated dependence
on $\theta$ of non-planar amplitudes at generic external momenta,
for which the Moyal phases associated with the interaction vertices 
do not factor out of Feynman diagram integrands, leading to
amplitudes that do not have a $\theta\to\infty$ limit uniform in
the external momenta. 

PIS theory is not a local quantum field theory. Beyond
the somewhat ``obvious'' non-locality (common to both the planar
and the non-planar sector) due to the
overall Moyal factors, there is also a non-locality which is 
associated with the vanishing of the total momenta entering
the holes of the non-planar amplitudes. 
As a consequence the effective action of PIS theory that
we will construct via the RG Wilson-Polchinski equation does
not have a functional integral representation based on 
some ``local'' (even in the non-commutative sense) space-time
action.  PIS theory represents an interesting example ---
and to our knowledge the first non-trivial one --- of a renormalizable 
theory of (partially integrated) Green functions
which can be rigorously defined and constructed 
only via the Wilson-Polchinski approach.

It is also intriguing to observe that PIS amplitudes are
in one-to-one correspondence, via the Eguchi-Kawai (EK) construction~\cite{ek},
with the multi-trace spherical amplitudes of a 0-dimensional
matrix model in the $N\to\infty$ limit. To see this, let us first briefly
recall the basic idea underlying the EK construction: the momenta $p_{ij}$
flowing through propagators of {\it planar } double line Feynman diagrams
of some (matrix or non-commutative) $d$-dimensional field theory
admit a representation in terms of {\it pseudo-momenta} $l_i$ as
$p_{ij} = l_i -l_j$, where the double indices $(i,j)$ label the propagator.
The pseudo-momenta $l_i$ for $i=1, \ldots, N$ are taken to form 
a regular lattice in momentum space centered around $p=0$ and of 
size equal to the ultra-violet
cut-off $\Lambda$. By replacing integrations over $d$-dimensional 
momenta $p$ with sums over the discrete indices $i$ one obtains
amplitudes which are regularized both in the UV and in the IR.
Then, a (regulated) planar Feynman diagram of (matrix
or non-commutative) $d$-dimensional field theory
equals a planar diagram of a 0-dimensional matrix model with
the same potential as the field theory and  with
propagator given by ${1\over N}{\delta_{ii^\prime}\,\delta_{j
j^\prime}\over (l_i -l_j)^2}$. It is maybe not widely appreciated that the map
between field theory and matrix model diagrams 
holds not only for the planar diagrams but more generally for 
PIS amplitudes: in fact this is precisely the property 
that {\it characterizes} such amplitudes. Thus the PIS restriction
appears to be very natural from the EK construction point of view:
the PIS sector of a (non-commutative) field theory is precisely 
the one described by the EK 0-dimensional matrix model.
In other words, the EK 0-dimensional matrix model is renormalizable
and captures the UV structure of non-commutative field theory.

The plan of this paper is the following: in Section 2 we write
the RG Wilson-Pol\-chinski equation for the large $\theta$ (lar\-ge $N$)
limit of non-com\-muta\-tive (matrix) field theory.
We use this equation to prove the re\-nor\-ma\-li\-zabi\-lity 
of the scalar 4-dimensional theory
and show that the marginal couplings also include the non-planar coupling 
$\sigma$ as\-socia\-ted with the 4-point functions with 2 holes and 2 legs
in each hole\footnote{The genus 0 RG equation of the scalar theory in 4d with
quartic interaction can be consistently projected
to the ``even'' parity sector: this consists of the amplitudes
with an {\it even} number of external legs in each hole.
In the explicit examples that we consider we will focus on this
sector.}. This is the coupling that in the matrix model corresponds to
the multi-trace operator $({\rm Tr} \phi^2)^2$.  In Section 3 we use the 
large $\theta$ RG equation to prove
the factorization property of PIS amplitudes and spell out its consequences
for the renormalization of both massive and massless non-commutative field
theories. In particular, in the massless case we derive the renormalized 
parametric equation that captures the anomalous dependence of planar 
amplitudes on the non-planar coupling $\sigma$ at $\Lambda=0$: we compute
at the lowest (2 loop) non-trivial order the generalized beta functions
that appear in this parametric equation. 
In Appendixes A and B we discuss the $\theta$ dependence of
spherical and higher-genus diagrams respectively.  We verify that
partially integrated amplitudes of genus $g$ go as $\theta^{-dg}$ for
$\theta\to\infty$, while amplitudes generic external momenta do not
have a uniform $\theta\to\infty$ limit.  In Appendix C we derive the
Wilson-Polchinski RG equation for the generating functional of
one-particle irreducible amplitudes in the large $N$ (large $\theta$)
limit.

\section{Wilson-Polchinski renormalization for\\ large $\theta$}

The  generating functional of connected amputated 
amplitudes of spherical topology for non-commutative field theory 
writes as
\be
H_\Lambda[\Omega]\!\equiv
\!\!\sum_{h=0}^{\infty}\,\sum_{k_1,\ldots, k_h}\int\!\prod_{i=1}^h
\prod_{\a_i=1}^{k_i}
\! dp^{\sss (i)}_{{\scriptscriptstyle \a_i}}
\,\d \Bigl(\sum_{\sss i,\,\a_i}
p^{\sss (i)}_{{\scriptscriptstyle \a_i}}\Bigr)\,
{\rm H}^{{\sss (h; \{ k_i\})}}_{\Lambda}[C_1,\ldots,C_h]\,\prod_{i=1}^{h}
\Omega_{k_i} (C_i)
\label{scfunctionalbis}
\ee
In the formula above 
${\rm H}^{{\sss (h; \{ k_i\})}}_{\Lambda}[C_1,\ldots,C_h]$ is
the connected amputated amplitude with $h$ holes labeled by the index
$i$, with $i=1,\ldots, h$. The $i$-th hole has $k_i$ external legs 
attached to it, whose momenta form the cyclically ordered set 
$C_i\equiv\{p^{\sss (i)}_{\sss 1},\ldots,p^{\sss (i)}_{\sss k_i}\}$.

In \cite{bgi} we proved that $H_\Lambda[\Omega]$ satisfies the following 
Wilson-Polchinski renormalization group equation
\bea
&&\!\!\!\!\!\!\Lambda\partial_\Lambda H_\Lambda \! =\! {1\over 2} \!\int\! dp\, \Ddot(p) 
\Biggl[\,\sum_{k,k^\prime} k k^\prime \!\!\int \!\! 
d p_1\ldots d p_{k-1}\, d q_1\ldots d q_{k^\prime-1}\times\\
&&\qquad\quad\times \, \Omega_{k+k^\prime-2}( p_1,\ldots, p_{k-1}, q_1,
\ldots, q_{k^\prime-1})\times \nonumber\\ 
&&\qquad\quad\times \, {\d H_\Lambda\over \d \Omega_k( p, p_1,\ldots,p_{k-1}) }\, 
{\d H_\Lambda \over \d \Omega_{k^\prime}(-p, q_1,\ldots, q_{k^\prime -1}) }+
\nonumber\\ 
&&\qquad\quad + 
\sum_k \sum_{i=1}^{k-1} k \int\!\! d p_1\ldots d p_{k-1}
\,\d\scl p_i-p\scr \times \nonumber\\
&&\qquad\times \, \Omega_{i-1}(p_1,\ldots, p_{i-1}) \,\Omega_{k-1-i}(p_{i+1},
\ldots, p_{k-1})\, 
{\d H_\Lambda \over \d \Omega_k( -p, p_1,\ldots, p_{k-1})  }\,\Biggr]\nonumber
\label{largenrg}
\eea
where $\Omega_0 \equiv 1$. In the equation above $\Ddot(p)\equiv \Lambda
\partial_\Lambda \Del(p)$, and $\Del(p)$ is the propagator, which is
regulated both by an ultra-violet cut-off $\Lambda_0$ and by an
infra-red one $\Lambda$.  Notice that $H_\Lambda$ depends on the
UV cut-off $\Lambda_0$ via the regulated propagators, though we will  
drop explicit reference to the ultra-violet scale $\Lambda_0$
in this section. 

\begin{figure}
\begin{center}
\includegraphics*[scale=.6, clip=false]{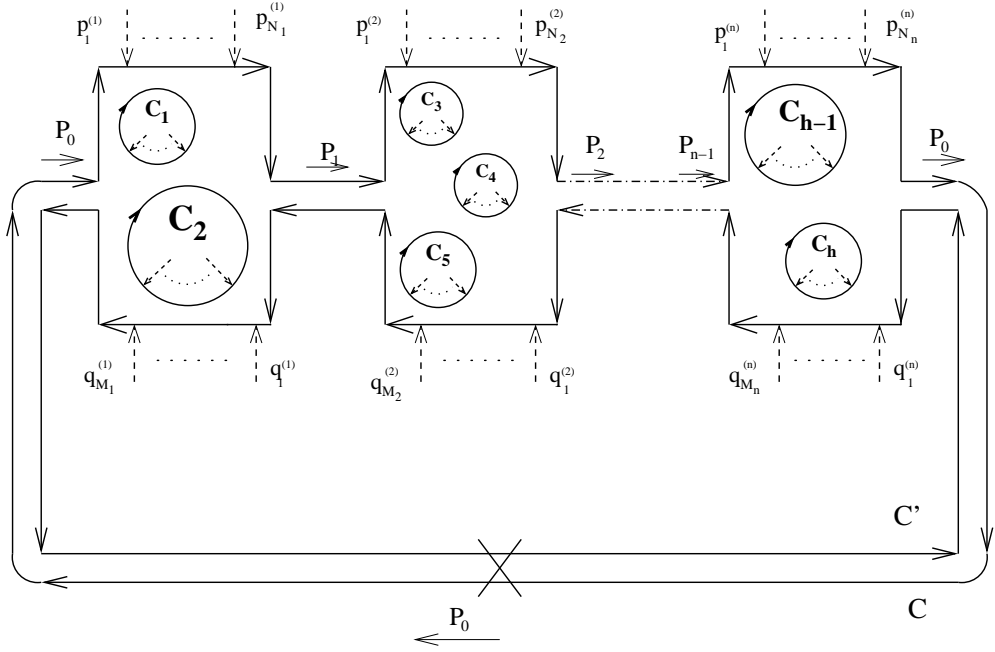}
\end{center}
\caption[x] {\footnotesize The RG equation for 1PI amplitudes. The crossed propagator gives the $\Ddot(P_0)$ factor. The dashed arrows are external lines}
\label{f:1PIfig}
\end{figure}

To analyse the renormalization properties of the non-commutative field
theory it is convenient to introduce the generating functional 
$\Gamma^\prime_\Lambda[\Omega]$ of the one-particle irreducible (1PI) 
spherical amplitudes  
$\Gamma^{\prime\,{\sss (h; \{ k_i\})}}_{\Lambda}[C_1,\ldots, C_h]$ :
\be
\Gamma^\prime_\Lambda[\Omega]\!\equiv
\!\!\sum_{h=0}^{\infty}\,\sum_{k_1,\ldots, k_h}\int\!\prod_{i=1}^h\Bigl[
\prod_{\a_i=1}^{k_i}\bigl[dp^{\sss (i)}_{{\scriptscriptstyle \a_i}}\bigr]\,
\Omega_{k_i} (C_i)\Bigr]
\,\d \Bigl(\sum_{\sss i,\,\a_i}p^{\sss (i)}_{{\scriptscriptstyle \a_i}}\Bigr)\,
{\Gamma}^{\prime\,{\sss (h; \{ k_i\})}}_{\Lambda}[C_1,\ldots,C_h]
\label{1PIfunctional}
\ee 
As proved in Appendix C, the RG equation for the 1PI functional writes as 
follows:
\bea
&&\Lambda\partial_\Lambda \Gamma_\Lambda = {1\over2}\,\sum_{n=1}^\infty
\int\! d{\rm P_0}\cdots d{\rm P_{n-1}}\,\Ddot({\rm P_0})\,\Del({\rm P_1})\cdots
\Del({\rm P_{n-1}})\times\nonumber\\
&&\qquad\times
\prod_{i=1}^n\Biggl[ \sum_{k_i}\sum_{I_i=0}^{k_i-2}\int\! \!\prod_{\a_i=1}^{I_i} 
\!\!dp^{(i)}_{\a_i} \!\!\prod_{\b_i=1}^{k_i-2-I_i}\!\! dq^{(i)}_{\b_i}\, k_i\, {\delta
\Gamma_\Lambda\over\delta\Omega_{k_i}({\rm P_{i-1}}, C_i,
{\scriptstyle -}{\rm P_{i}}, C^\prime_i)}\Biggr]\times\nonumber\\
&&\qquad\times\, 
\Omega_{\textstyle{\sss\sum_i I_i}}(C_n,\ldots, C_1)\,
 \Omega_{\textstyle{\sss\sum_i k_i-2-I_i}}(C^\prime_1,\ldots, C^\prime_n)
\label{1PIlargenrg}
\eea
where $C_i\equiv \{p^{(i)}_{\a_i}\}$, $C^\prime_i\equiv \{q^{(i)}_{\b_i}\}$
and $\Gamma_\Lambda[\Omega]$ is defined by 
\be
\Gamma_\Lambda[\Omega] = \Gamma^\prime_\Lambda[\Omega] +{1\over2}\!\int\!\!dp 
\,\Dinv(p)\, \Omega_2(p,-p)
\ee  
Eq. (\ref{1PIlargenrg}) translates into evolution equations for
the amplitudes $\Gamma^{(h)}_{\Lambda}$ which
have following schematic structure
\bea
&&\!\!\!\!\!\!\!\!\Lambda\partial_\Lambda \Gamma^{(h+2)}_\Lambda
[C_1,\ldots,C_{h+2}]= {1\over2}
 \int\! d{\rm P_0}\,\Ddot({\rm P_0}) \sum_{n=1}^\infty
\mathop{{\sum}^\prime}\,\Del({\rm P_1})\cdots
\Del({\rm P_{n-1}})\times\nonumber\\
&&\qquad\quad\times\prod_{i=1}^n
\Gamma^{(h_i+1)}_\Lambda\Bigl[ C_{I^{(i)}_1},\ldots, C_{I^{(i)}_{h_i}},
\{{\rm P_{i-1}}, p^{(i)}_1,\ldots, p^{(i)}_{N_i}, {{\sss -}\rm P_i},
q^{(i)}_1,\ldots, q^{(i)}_{M_i}\}\Bigr]\nonumber\\
\label{1PImess}
\eea
and are graphically represented in Figure~\ref{f:1PIfig}.
The R.H.S. of this equation involves several sums which we indicated with 
$\mathop{{\sum}^\prime}$: (a) 
the sum over  the possible ways
to select 2 holes $C$ and $C^\prime$ 
(with $N$ and $M$ external legs respectively) among the $h+2$ holes 
$C_1,\ldots, C_{h+2}$ of 
the amplitude on the L.H.S.; (b) the sum over the ways to partition 
the external
momenta of $C$ and $C^\prime$ into $n$ subsets of consecutive momenta 
$\{p^{(i)}_1,\ldots, p^{(i)}_{N_i}\}$
and $\{q^{(i)}_1,\ldots, q^{(i)}_{M_i}\}$, with $\sum_{i=1}^n N_i= N$ 
and $\sum_{i=1}^n M_i =M$; (c) the sum over the possible ways to distribute
the remaining $h$ holes into the $n$ sets denoted in Eq. (\ref{1PImess}) by
$\{C_{I^{(i)}_1},\ldots, C_{I^{(i)}_{h_i}}\}$, with $\sum_{i=1}^n h_i =h$.
The momenta $P_i$, with $i=1,\ldots,n-1$, are functions of the loop
momentum $P_0$ and of the external momenta defined by the relations
\be
{\rm P_i} = {\rm P}_{i-1} + \sum_{\a_i=1}^{N_i} p^{(i)}_{\a_i} +
\sum_{\b_i=1}^{M_i} q^{(i)}_{\b_i} + \sum_{\gamma_i=1}^{h_i}\,{\rm P}
\bigl(C_{I^{(i)}_{\gamma_i}}\bigr)
\label{conservation}
\ee
where ${\rm P}(C)$  is the total momentum entering the hole $C$.
Note that, thanks to Eq. (\ref{conservation}),  
the total momenta entering the holes of the amplitudes
$\Gamma^{(h_i+1)}$ appearing on the R.H.S. of Eq. (\ref{1PImess})
are linear combinations of the momenta ${\rm P}(C_i)$ entering the holes
of the amplitude $\Gamma^{(h+2)}$ on the L.H.S. Thus,
if the momenta configurations appearing in the L.H.S. are  {\it exceptional} 
--- i.e. if all the ${\rm P}(C_i)=0$ ---
then the amplitudes involved in the R.H.S. are also 
evaluated at exceptional momenta. We will call these amplitudes
{\it partially integrated spherical} (PIS) amplitudes.
Hence, Eq. (\ref{1PImess}) implies that the RG evolution can be consistently
restricted to PIS amplitudes. 

It is worth remarking that Eq.~(\ref{1PImess}) predicts the factorization
of the $\theta$
dependence of PIS amplitudes that we anticipated in the Introduction
and worked out in the Appendix A.
Indeed assume that at {\it any} scale $\Lambda_0$ the $\theta$ dependence
of a PIS amplitude with $h$ holes 
$C_1,\ldots C_h$ is the product
of the Moyal factors ${\rm e}^{-i\Phi(C_i)}$
associated with each hole: then it can be verified
that the product of the Moyal factors of the
amplitudes $\Gamma^{\sss (h_i+1)}$ which enter the
R.H.S. of the evolution equation (\ref{1PImess}) equals the product 
of the Moyal factors associated
with the $h+2$ holes $C_1,\ldots,C_{h+2}$ which appear on the L.H.S. 
In other words, the RG evolution equation for large $\theta$ implies 
that the $\theta$ dependence of PIS amplitudes is restricted
to the Moyal factors and hence $\theta$ does not run.

Let us therefore introduce the generating functional for PIS amplitudes:
\be
\!\!{\Gammapis}_\Lambda [\Omega ]\!\equiv
\sum_{h=0}^{\infty}\,\Lambda^{d(h - 1)}\!\!\!\!\sum_{k_1,\ldots, k_h}\!
\!\int\!\prod_{i=1}^h
\Bigl[\prod_{\a_i=1}^{k_i}\!
\bigl[dp^{\sss (i)}_{{\scriptscriptstyle \a_i}}\bigr]\,
\Omega_{k_i} (C_i)\,\d \bigl({\rm P}(C_i)\bigr)\Bigr]\,
{\Gamma}^{{\sss (h; \{ k_i\})}}_{\Lambda}[C_1,\ldots,C_h]\
\label{PIS1PI}
\ee
where the $\Lambda^{d(h - 1)}$ factors have been introduced to keep
$\Gammapis_\Lambda$ dimensionless. $\Gammapis_\Lambda$  satisfies an
RG evolution equation which is only slightly different than Eq. 
(\ref{1PIlargenrg}):
\bea
&&\!\!\!\!\!\!\!\! {\cal D}_\Lambda{\Gammapis}_\Lambda = 
{1\over2}\,\Lambda^d\sum_{n=1}^\infty
\int\! d{\rm P_0}\cdots d{\rm P_n}\,\Ddot({\rm P_0})\,\Del({\rm P_1})\cdots
\Del({\rm P_{n-1}})\times\nonumber\\
&&\quad\times
\prod_{i=1}^n\Biggl[\sum_{k_i}\sum_{I_i=0}^{k_i-2}
\int\! \!\prod_{\a_i=1}^{I_i} 
\!\!dp^{(i)}_{\a_i} \!\!\prod_{\b_i=1}^{k_i-2-I_i}\!\! dq^{(i)}_{\b_i}\, k_i\, {\delta
\Gammapis_\Lambda\over\delta\Omega_{k_i}({\rm P_{i-1}},\{p^{(i)}_{\a_i}\},
{\scriptstyle -}{\rm P_{i}},\{q^{(i)}_{\b_i}\})}\Biggr]\times\nonumber\\
&&\qquad\qquad\times\, \delta(\sum_{i,\,\a_i}p^{(i)}_{\a_i})\,\Omega_{\textstyle{\sss\sum_i I_i}}(\{p^{(n)}_{\a_n}\},\ldots,\{p^{(1)}_{\a_1}\})\times\nonumber\\
&&\qquad\qquad\qquad \times\, \delta(\sum_{i,\,\b_i}q^{(i)}_{\b_i})
\,\Omega_{\textstyle{\sss\sum_i k_i-2-I_i}}(\{q^{(1)}_{\b_1}\},\ldots,\{q^{(n)}_{\b_n}\})
\label{PISlargenrg}
\eea
where
\be
{\cal D}_\Lambda \equiv \Lambda \partial_\Lambda + d- d\sum_k\!\int\! 
\prod_{i=1}^k\,d p_i\,\Omega_k(p_1,\ldots,p_k)\,{\delta\over\delta
\Omega_k(p_1,\ldots,p_k)}
\ee
and we adopted the convention that 
\be
\delta({\rm P}(C))\Omega_k(C)\to \Lambda^{-d}\qquad\quad{\rm for}\,\, k=0
\ee

Starting from Eq. (\ref{PISlargenrg}) one can prove the renormalizability
of the theory of PIS amplitudes in the Wilson-Polchinski sense. 
The evolution equation determines $\Gammapis_\Lambda$
at an arbitrary value of $\Lambda$ once initial conditions are
chosen. Initial conditions for the couplings 
are chosen either at the UV ``high'' scale $\Lambda_0$ --- for the
irrelevant couplings --- or at the ``low'' scale
$\Lambda_R$ --- for the marginal and relevant ones. 
Renormalizability is proven by showing that the functional
${\Gammapis}_{\Lambda,\, \Lambda_0}$ determined by these
initial conditions has a finite limit as $\Lambda_0\to\infty$ ---
as marginal and relevant couplings at the scale $\Lambda_R$ are
kept fixed.

The proof of renormalizability of the (non-local) theory of PIS amplitudes
follows the same arguments \cite{polchinski}, based  on dimensional
analysis, which apply to (local) commutative
theories. Eq. (\ref{1PImess}) shows that the dependence of
the amplitudes on $\Lambda$, for $\Lambda$ much larger than
the external momenta, is the same as in the commutative case,
that is 
\be
\Gamma_\Lambda^{{\sss (h; \{ k_i\})}}[C_1,\ldots,C_h]\sim \Lambda^{d -\sum_{i=1}^h n(C_i)}
\label{dimension}
\ee
where $n(C_i)$ is the number of legs attached to the hole $C_i$,
and where possible logarithmic dependence is not explicitly
indicated. Indeed, if we choose for concreteness a sharp cut-off
for the propagator\footnote{Any momentum cut-off which falls off 
sufficiently fast will do.}:
\be
\Del(p) = {1\over p^2 +m^2}\,\bigl[\Theta(\Lambda^2- p^2)- \Theta(\Lambda^2_0 - p^2)\bigr]
\ee
we have 
\be
\Ddot(P_0) \sim \delta(\Lambda^2 - P_0^2)
\ee
It is then immediate to verify that the scaling law (\ref{dimension})
is consistent with the evolution equation (\ref{1PImess}). 
Specializing now our considerations to the $d=4$ case, it follows
from Eq. (\ref{dimension}) that
the relevant and marginal couplings are those associated
with amplitudes with 2 or 4 external legs:
\bea
&&{\rm N}_{\sss m^2}[\Gammapis_\Lambda]\equiv 
\Gamma_\Lambda^{\sss \{2\}}(0,0)\nonumber\\
&&{\rm N}_{\sss Z}[\Gammapis_\Lambda]\equiv 
\partial_{p^2}\,\Gamma_\Lambda^{\sss \{2\}}(p,-p)\Big|_{p=0}\nonumber\\
&&{\rm N}_{\sss g}[\Gammapis_\Lambda]\equiv 
\Gamma_\Lambda^{\sss \{4\}}(0,0,0,0)\nonumber\\
&&{\rm N}_{\sss \sigma}[\Gammapis_\Lambda]\equiv 
\Gamma_\Lambda^{\sss \{2,2\}}(0,0; 0,0)
\label{runningcouplings}
\eea
where $\Gamma_\Lambda^{\sss \{2\}}$, 
$\Gamma_\Lambda^{\sss \{4\}}$ are the planar 2- and 4-point functions
and $\Gamma_\Lambda^{\sss \{2,2\}}$ is the {\it non-planar} 4-point function
with 2 holes. The renormalization conditions for the massive
theory are set at a low energy
scale $\Lambda_R$:
\bea
&{\rm N}_{\sss m^2}[\Gammapis_\Lambda]\Big|_{\Lambda=\Lambda_R} = 0\qquad&
{\rm N}_{\sss Z}[\Gammapis_\Lambda]\Big|_{\Lambda=\Lambda_R}= 
Z-1\nonumber\\
&{\rm N}_{\sss g}[\Gammapis_\Lambda]\Big|_{\Lambda=\Lambda_R} = g\qquad
&{\rm N}_{\sss \sigma}[\Gammapis_\Lambda]\Big|_{\Lambda=\Lambda_R}=\sigma
\label{renormalizationconditions}
\eea
For the massless theory, for which $m^2=0$, one must replace
the first of the equations above with
\be
{\rm N}_{\sss m^2=0}[\Gammapis_\Lambda]\Big|_{\Lambda= 0}= 0
\label{renconditionsmassless}
\ee
and keep the others unchanged. All the other couplings are irrelevant
and thus can be chosen arbitrarily at the UV scale $\Lambda_0$.
In the Wilson-Polchinski approach the non-commutative parameter $\theta$
appears in the initial condition for $\Gammapis_\Lambda$.  
The non-commutative Moyal theory is defined by 
setting the $\theta$ dependence  of the irrelevant couplings 
with  $h$ holes, at the scale $\Lambda_0$, 
to be the product of the Moyal factors associated with the same $h$ holes:
as we remarked above, the RG evolution equation ensures that the
$\theta$ dependence is preserved by the renormalization flow.

By integrating the evolution equation (\ref{PISlargenrg}) with the boundary
conditions (\ref{renormalizationconditions}) 
(or (\ref{renconditionsmassless})) and using the scaling property 
(\ref{dimension}) one shows, as in the usual
Polchinski framework, that the amplitudes have a finite limit
for $\Lambda_0\to\infty$. The same argument shows that 
that amplitudes evaluated at low momenta $p$ and scale $\Lambda$
depend on the values of the irrelevant couplings at the scale
$\Lambda_0$ as positive powers of $\Lambda/\Lambda_0$ or
$p/\Lambda_0$. 

Let us comment on the relevance of our results to
the celebrated IR-UV problem of non-commutative field
theories. In the approaches to renormalization of non-commutative
theories proposed so far~\cite{seiberg, griguolo} one introduces
counterterms only for planar divergences: non-planar
divergences are regulated by the effective UV cut-off 
$1/(\theta p)$, where $p$ is the momentum entering a hole
of the diagram. Therefore, if $p$ is external, the amplitude
develops an IR/UV divergence as $p\to 0$. 
However, as stressed in the Introduction, non-planar divergences also 
occur as sub-divergences of
planar amplitudes: the consequence of this is that
even the {\it planar} sector of the theory is not correctly
renormalized if only planar counterterms are introduced.
The Wilson-Polchinski approach makes this evident, 
since, as we have already emphasized, the RG equation
inevitably couples planar and non-planar amplitudes.
In the following subsection we will show explicitly that, if only planar
counterterms are introduced, one can remove the UV divergences of the
planar Wilsonian action at a given renormalization scale $\Lambda=\Lambda_R$,
but not at {\it all} scales $\Lambda$: this is precisely the manifestation 
at the planar level of the IR/UV problem which plagues non-planar
amplitudes renormalized according to planar renormalization.

Our theory, on the other hand, includes counterterms associated
with both planar and non-planar couplings: in the 4 dimensional
scalar case, for example, one must introduce a non-planar counterterm
associated with the $\sigma$ coupling. This non-planar counterterm, 
which corresponds to the term 
\be
\Lambda^4\,\sigma \Bigl[\int\!\! dp\, \Omega_{2}(\{p, {\sss -}p\})\Bigr]^2
\label{sigmacounterterm}
\ee
of the effective action in Eq. (\ref{PIS1PI}), is evidently
non-local.  We will see that such a non-local counterterm  
is essential for the UV finiteness of all the amplitudes, both 
planar and non-planar, at {\it any} value of the
floating Polchinski cut-off $\Lambda$: maybe surprisingly, the
non-local counterterm  (\ref{sigmacounterterm})
cancels ``local'' (in the non-commutative sense) 
divergences of the planar part of the effective action.
We will see an explicit example of this mechanism
in the next subsection where we compute the 2-point planar amplitude
$\Gamma_\Lambda^{\sss \{2\}}(p,-p)$ at 2 loops.

Of course, had we not restricted ourselves to exceptional momenta,
counterterms required to remove non-planar divergences should have had
a complicated dependence on $\theta$ and the external momenta and thus
a very non-local space-time structure. Fortunately, as we explained above,
non-planar sub-divergences occurs only at exceptional momenta: it is
this that makes possible the removal of all UV divergences of 
PIS amplitudes by means of counterterms that have a simple and
``universal'' $\theta$ and momentum dependence: their  non-locality 
reduces to the product of the Moyal factors and
momentum delta functions associated with each hole.

\subsection{Non-planar sub-divergences of planar amplitudes: 
a two-loop example}

The ``bare'' 2-point function computed at two loops is
\bea
&&\!\!\!\!
\Gamma^{{\sss \{2\}}}_{\Lambda,\,\Lambda_0}(p,-p) = {\delta m^2\over 2} + 
{p^2\over 2}\,\delta z + \Bigl(g_0 + {\sigma_0\over 2}\Bigr)\,
I^{(1)}_{\Lambda,\,\Lambda_0} + \delta m^2\,\Bigl(g_0 + {\sigma_0\over 2}
\Bigr)\,
I^{(2)}_{\Lambda,\,\Lambda_0}(0)\nonumber\\
&&\!\!\qquad\qquad\qquad +\, {g_0^2\over 2}\,
I^{(3)}_{\Lambda,\,\Lambda_0}(p) + 2\,\Bigl(g_0 + {\sigma_0\over 2}
\Bigr)^2 \, 
I^{(2)}_{\Lambda,\,\Lambda_0}(0)\,I^{(1)}_{\Lambda,\,\Lambda_0} + O(\hbar^3)
\label{bare2point}
\eea
where $I^{(i)}_{\Lambda,\,\Lambda_0}$, with $i=1,\ldots,3$, are the following 
IR and UV regulated Feynman integrals
\bea
&&I^{(1)}_{\Lambda,\,\Lambda_0}\equiv\int\!{d k\over (2\pi)^4}\,
\Delta_{\Lambda,\,\Lambda_0}(k)\nonumber\\
&&I^{(2)}_{\Lambda,\,\Lambda_0}(p)\equiv\int\!{d k\over (2\pi)^4}\,
\Delta_{\Lambda,\,\Lambda_0}(k)\,\Delta_{\Lambda,\,\Lambda_0}(p-k)\nonumber\\
&&I^{(3)}_{\Lambda,\,\Lambda_0}(p)\equiv\int\!{d k\over (2\pi)^4}\,
{d q\over (2\pi)^4}\,
\Delta_{\Lambda,\,\Lambda_0}(k)\,\Delta_{\Lambda,\,\Lambda_0}(q)\,
\Delta_{\Lambda,\,\Lambda_0}(p-k-q)
\eea
The bare couplings $\delta m^2\equiv m_0^2 -m^2$, $\delta z\equiv Z_0 - 1$, 
$g_0$ and $\sigma_0$, defined by the equations
\bea
&{\rm N}_{\sss m^2}[\Gammapis_\Lambda]\Big|_{\Lambda=\Lambda_0}=
\delta m^2
&\qquad{\rm N}_{\sss Z}[\Gammapis_\Lambda]\Big|_{\Lambda=\Lambda_0}= 
\delta z\nonumber\\
&{\rm N}_{\sss g}[\Gammapis_\Lambda]\Big|_{\Lambda=\Lambda_0}= g_0
&\qquad{\rm N}_{\sss \sigma}[\Gammapis_\Lambda]\Big|_{\Lambda=\Lambda_0}
=\sigma_0
\label{barecouplings}
\eea
are functions of the renormalized ones $m^2$, $g$ and $\sigma$ 
determined by
the renormalization conditions (choosing $Z=1$).  
Let us restrict ourselves to the massive
case: Eqs. (\ref{renormalizationconditions}) give
\bea
&&\!\!\!\!{\delta m^2\over 2} = -\, \Bigl(g + {\sigma\over2}\Bigr)\, 
I^{(1)}_{\Lambda_R,\,\Lambda_0} 
- g^2\, \Bigl[{1\over 2}\,
I^{(3)}_{\Lambda_R,\,\Lambda_0}(0) - {3\over 2}\,
I^{(2)}_{\Lambda_R,\,\Lambda_0}(0)\,I^{(1)}_{\Lambda_R,\,\Lambda_0} \Bigr] +
\nonumber\\
&&\qquad\qquad +\, 2\,\Bigl(g + {\sigma\over 2}\Bigr)^2\,
I^{(2)}_{\Lambda_R,\,\Lambda_0}(0)\,I^{(1)}_{\Lambda_R,\,\Lambda_0}+ O(\hbar^3)
\nonumber\\
&&\!\!\!\!{\delta z\over 2} = - {g^2\over 2}\, \partial_{p^2}\,
I^{(3)}_{\Lambda_R,\,\Lambda_0}(p)\Big|_{p^2=0} + O(\hbar^3)\nonumber\\
&&\!\!\!\!g_0 = g - 2\,g^2\,I^{(2)}_{\Lambda_R,\,\Lambda_0}(0) + O(\hbar^2)
\nonumber\\
&&\!\!\!\!\sigma_0 = \sigma - (3\,g^2 + 4\, g\,\sigma + \sigma^2)\,I^{(2)}_{\Lambda_R,\,\Lambda_0}(0) + O(\hbar^2)
\label{2loopsbarecouplings}
\eea
Substituting now Eqs. (\ref{2loopsbarecouplings}) in Eq. (\ref{bare2point}), we
compute the renormalized 2-point function:
\bea
&&\!\!\!\!\!\!\!\!\!\!
\Gamma^{{\sss \{2\}}}_{\Lambda,\,\Lambda_0;\,\Lambda_R}(p,-p) = 
\Bigl(g + 
{\sigma\over 2}\Bigr)\,\Bigl[I^{(1)}_{\Lambda,\,\Lambda_0}-
I^{(1)}_{\Lambda_R,\,\Lambda_0}\Bigr]+\nonumber\\
&&\qquad +\, g^2 \Bigl[{1\over 2}\,
I^{(3)}_{\Lambda,\,\Lambda_0}(p) - {1\over 2}\, 
I^{(3)}_{\Lambda_R,\,\Lambda_0}(0) - {1\over 2}\,p^2\, \partial_{p^2}\,I^{(3)}_{\Lambda_R,\,\Lambda_0}(0)\Bigr] +\nonumber\\ 
&&\qquad -
{3\over 2}\,g^2\, I^{(2)}_{\Lambda_R,\,\Lambda_0}(0)\, \Bigl[
I^{(1)}_{\Lambda,\,\Lambda_0}-I^{(1)}_{\Lambda_R,\,\Lambda_0}\Bigr] 
+ \nonumber\\
&&\qquad +\,2\,\Bigl(g + {\sigma\over 2}\Bigr)^2\,\Bigl[
I^{(2)}_{\Lambda,\,\Lambda_0}(0)- I^{(2)}_{\Lambda_R,\,\Lambda_0}(0)\Bigr]\,
\Bigl[I^{(1)}_{\Lambda,\,\Lambda_0} - I^{(1)}_{\Lambda_R,\,\Lambda_0}\Bigr] +\,  O(\hbar^3)\nonumber\\
&&\qquad = {g^2\over 2}\, \Bigl[I^{(3)}_{\Lambda,\,\Lambda_0}(p) -  
I^{(3)}_{\Lambda_R,\,\Lambda_0}(0)-\,p^2\, \partial_{p^2}\,I^{(3)}_{\Lambda_R,\,\Lambda_0}(0) + 
3\,I^{(2)}_{\Lambda_R,\,\Lambda_0}(0)\,I^{(1)}_{\Lambda_R,\,\Lambda}\Bigr] +
\nonumber\\
&&\qquad -\,\Bigl(g +
{\sigma\over 2}\Bigr)\,I^{(1)}_{\Lambda_R,\,\Lambda}
+ 2\,\Bigl(g + {\sigma\over 2}\Bigr)^2\,
I^{(2)}_{\Lambda_R,\,\Lambda}(0)\,
I^{(1)}_{\Lambda_R,\,\Lambda}+\,  O(\hbar^3)
\label{2pointsrenormalized}
\eea
where we used the fact that $I^{(1)}_{\Lambda_R,\,\Lambda_0}=
I^{(1)}_{\Lambda_R,\,\Lambda}+I^{(1)}_{\Lambda,\,\Lambda_0}$
and $I^{(2)}_{\Lambda_R,\,\Lambda_0}(0)=
I^{(2)}_{\Lambda_R,\,\Lambda}(0)+I^{(2)}_{\Lambda,\,\Lambda_0}(0)$.
From  Eq. (\ref{2pointsrenormalized}) we obtain the following expression
for the ``running'' mass coupling
\bea
&&\!\!\!\!\!\!\!{\rm N}_{\sss m^2}[\Gammapis_\Lambda] = 
{g^2\over 2}\, \Bigl[I^{(3)}_{\Lambda,\,\Lambda_0}(0) -  
I^{(3)}_{\Lambda_R,\,\Lambda_0}(0)+ 
3\,I^{(2)}_{\Lambda_R,\,\Lambda_0}(0)\,I^{(1)}_{\Lambda_R,\,\Lambda}\Bigr] +
\nonumber\\
&&\qquad\qquad -\,\Bigl(g +
{\sigma\over 2}\Bigr)\,I^{(1)}_{\Lambda_R,\,\Lambda} + 
2\,\Bigl(g + {\sigma\over 2}\Bigr)^2\,
I^{(2)}_{\Lambda_R,\,\Lambda}(0)\,
I^{(1)}_{\Lambda_R,\,\Lambda} +\,  O(\hbar^3)
\label{runningmass}
\eea

\begin{figure}
\begin{center}
\includegraphics*[scale=.4, clip=false]{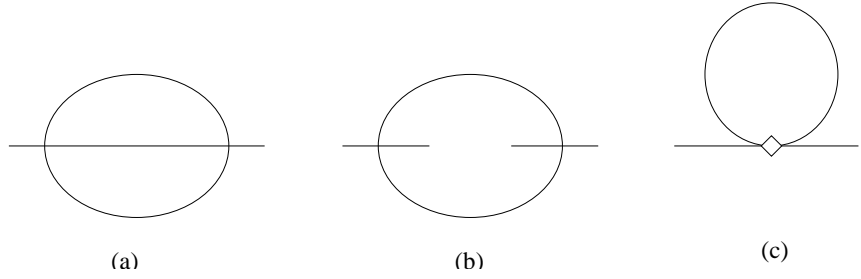}
\end{center}
\caption[x] {\footnotesize A planar 2-loop diagram (a), with a non-planar
divergent subgraph (b). (c) is the 1-loop correction to $\Gamma^{\sss \{2\}}$ 
with the 1-loop $\sigma$ correction counterterm. (c) 
cancels non-planar sub-divergences like (b)}
\label{f:diagrams2}
\end{figure}
Since the terms in the last two lines  of Eq. (\ref{runningmass}) are
$\Lambda_0$ independent, UV finiteness of the coupling 
${\rm N}_{\sss m^2}[\Gammapis_\Lambda]$ relies on
the UV finiteness of the expression in square brackets. 
Note the following: the last (divergent) term of this expression --- 
$3\,I^{(2)}_{\Lambda_R,\,\Lambda_0}(0)\,I^{(1)}_{\Lambda_R,\,\Lambda}$ --- 
originates from the 1-loop contribution to the $\sigma_0$ counterterm. This 
term cancels non-planar sub-divergences of the 2-loop amplitude, 
like the non-planar sub-divergence of $I^{(3)}$ shown in 
Figure~\ref{f:diagrams2}~(b).
Had we not included the non-planar coupling $\sigma$ among
the marginal ones this counterterm would be absent, and, as we will
see temporarily, the amplitude (\ref{runningmass}) would be UV divergent
for $\Lambda \not= \Lambda_R$. 
This shows explicitly that non-planar divergences  --- at exceptional
momenta --- also affect the UV behavior of planar amplitudes at higher loops.
To show the UV finiteness of ${\rm N}_{\sss m^2}[\Gammapis_\Lambda]$ let us
consider the identity:
\bea
&&\!\!\!\!\!\!\!\!\!\!I^{(3)}_{\Lambda,\,\Lambda_0}(0) -  
I^{(3)}_{\Lambda_R,\,\Lambda_0}(0)
=-3 \!\int\!
{d k\over (2\pi)^4}\,
\Delta_{\Lambda_R,\,\Lambda}(k)\,I^{(2)}_{\Lambda,\,\Lambda_0}(k)\,+\nonumber\\
&&\qquad -\, I^{(3)}_{\Lambda_R,\,\Lambda}\!\!- 3\!\int\!{d k\over (2\pi)^4}\,
{d q\over (2\pi)^4}\,
\Delta_{\Lambda_R,\,\Lambda}(k)\,\Delta_{\Lambda_R,\,\Lambda}(q)\,
\Delta_{\Lambda,\,\Lambda_0}(k+q)
\label{divergence2}
\eea
The only UV divergent term in the R.H.S. of the equation above
is the integral in the first line, which can be written as
\be
-3 \!\int\!
{d k\over (2\pi)^4}\,
\Delta_{\Lambda_R,\,\Lambda}(k)\,I^{(2)}_{\Lambda,\,\Lambda_0}(k) 
= -3\, I^{(1)}_{\Lambda_R,\,\Lambda}\,I^{(2)}_{\Lambda,\,\Lambda_0}(0)+
R (\Lambda_R,\,\Lambda,\,\Lambda_0)
\ee
where $R (\Lambda_R,\,\Lambda,\,\Lambda_0)$ is finite as $\Lambda_0\to\infty$.
The non-planar counterterm
$3\,I^{(2)}_{\Lambda_R,\,\Lambda_0}(0)\,I^{(1)}_{\Lambda_R,\,\Lambda}$
in Eq. (\ref{runningmass}) is required precisely to cancel the divergence
in Eq. (\ref{divergence2}).

Note that the non-planar counterterm $3\,I^{(2)}_{\Lambda_R,\,\Lambda_0}(0)\,I^{(1)}_{\Lambda_R,\,\Lambda}$ vanishes at $\Lambda=\Lambda_R$. This means that
the 2-point renormalized amplitude $\Gamma_\Lambda^{\sss \{2\}}(p)$
would be UV finite {\it at $\Lambda=\Lambda_R$} also
in a renormalization framework that did not include $\sigma$ 
among the marginal couplings \cite{seiberg}.
The fact that in such a framework the amplitudes at $\Lambda=\Lambda_R$
are UV finite while the Wilsonian running couplings (like 
${\rm N}_{\sss m^2}[\Gammapis_\Lambda]$) are not,
is the manifestation of the IR/UV difficulty
which occurs when one does not take into account
non-planar divergences. In the next Section we will generalize
this observation by showing that {\it planar}
renormalized amplitudes of the {\it massive theory} evaluated 
at $\Lambda =\Lambda_R$  
do not depend on the renormalized coupling $\sigma$. This indeed implies
that one can compute planar amplitudes of the massive theory
at $\Lambda=\Lambda_R$ forgetting about the non-planar marginal 
coupling $\sigma$. We will also show that this is {\it not} true
for the massless theory, whose planar amplitudes at $\Lambda=0$
have an ``anomalous'' dependence on $\sigma$ --- this is
how the interplay between the IR and the UV manifests
itself in our theory, which, nevertheless, is both renormalizable in 
the Wilsonian sense and completely free of IR/UV divergences.

\section{The Parametric Equation}

We have seen that renormalization of PIS amplitudes 
requires including the non-planar coupling
$\sigma$ among the relevant couplings. This makes all the
amplitudes  --- both the planar and the non-planar at exceptional
momenta --- finite. In this Section we will show that,
although the renormalized PIS amplitudes depend
on {\it four} relevant couplings 
--- $m^2$, $Z$, $g$ and $\sigma$ ---
the {\it planar} sector of the theory, at $\Lambda=\Lambda_R$,  
is controlled only by {\it
three} (suitable combinations) of them. Therefore, in a sense, the non-planar
coupling $\sigma$ can be thought of as an {\it evanescent} coupling
of the planar theory: the corresponding counterterm is required
to make planar amplitudes UV finite at any scale $\Lambda$, but renormalized 
planar amplitudes are at $\Lambda=\Lambda_R$, essentially, 
independent of the renormalized value of $\sigma$. To be more
precise, we will see that the latter statement is literally correct
only for the massive theory. 
In the massless theory the planar renormalized amplitudes, 
evaluated at the scale $\Lambda=0$, {\it do} 
depend on $\sigma$: however they satisfy a differential
equation of first order in the derivatives with respect to
the renormalized couplings. This implies that they are
independent of a certain combination
of $\sigma$ and $g$. In all cases the planar theory has one
less marginal parameter than the full (PIS) theory.

To show this point, we will start by recalling 
that the RG evolution equation in the large $\theta$ (or, in the
case of matrix theories, large $N$) limit
is an equation that, 
unlike the ordinary Wilson-Polchinski
RG equation,  is of {\it first order} in the derivatives of the 
generating functional with respect to the sources $\Omega_k$:
it can therefore be written in the form
\be
\Lambda\partial_\Lambda\,\Gammapis_\Lambda = R\Bigl[\Omega_k, 
{\delta\Gammapis_\Lambda\over \delta\Omega_k}\Bigr]\equiv
R\Bigl[\Omega_k, \Gammapis_{\Lambda , \Omega_k}\Bigr]
\ee 
where $R$ is the functional of the sources and the first order derivatives
of $\Gammapis_\Lambda$ that appears in the R.H.S. of Eq. (\ref{PISlargenrg}).
Suppose now that the generating 
functional $\Gammapis_\Lambda$
satisfies at the scale $\Lambda=\Lambda_0$ a differential equation
of the form
\be
E[{\partial_{\rho_0^{\sss (a)}}\Gammapis_{\Lambda_0}}]= 0
\label{nonlinear}
\ee
where $E$ is a --- not necessarily linear --- function of the first order 
derivatives 
$\Gammapis_{\Lambda}^{\sss (a)}\equiv {\partial_{\rho_0^{\sss (a)}}}\,\Gammapis_\Lambda$
of $\Gammapis_{\Lambda}$ with respect to the (bare) coupling constants  
$\rho_0^{(a)}$ (with $a$ running over the set  
$\{m^2, Z, g,\sigma\}$). It is important that
$E$ does not depend explicitly on the sources $\Omega_k$.
Then
\be
\Lambda\partial_\Lambda\,E[\Gammapis_{\Lambda}^{\sss (a)}]
= {\partial E\over\partial \Gammapis_{\Lambda}^{\sss (a)}}\,
\Lambda\partial_\Lambda\,\Gammapis_{\Lambda}^{\sss (a)}=
 {\partial E\over\partial \Gammapis_{\Lambda}^{\sss (a)}}\,{\delta\,R\over
\delta\Gammapis_{\Lambda ,\Omega_k}}\,
{\delta \Gammapis_{\Lambda}^{\sss (a)}\over\delta \Omega_k}= {\delta\,R\over
\delta\Gammapis_{\Lambda ,{\sss \Omega_k}}}\,  {\delta 
E\over\delta \Omega_k}
\label{parametric1}
\ee
This equation shows that if  $E=0 $ for $\Lambda=\Lambda_0$, $E=0$ identically
in $\Lambda$. In particular, we can choose $E$ as follows 
\be
E = \partial_{\sigma_0}\Gammapis_\Lambda - 
\Bigl(\partial_{m_0^2}\Gammapis_\Lambda \Bigr)^2
\ee
Since at the scale $\Lambda_0$ $E=0$ by definition, it follows that 
\be
 \partial_{\sigma_0}\Gammapis_\Lambda - 
\Bigl(\partial_{m_0^2}\Gammapis_\Lambda \Bigr)^2= 0
\label{parametric2}
\ee
for any $\Lambda$. Eq. (\ref{parametric2}) is the analogue for PIS theory
of the celebrated factorization property of large $N$ matrix models
\be
\bigl\langle \bigl({\rm Tr} \Phi^k\bigr)^2 \bigr\rangle =\bigl(
\bigl\langle {\rm Tr} \Phi^k\bigr\rangle\bigr)^2 
\ee
From the previous discussion it is apparent that factorization
is a direct consequence --- in the Wilson-Polchinski framework ---
of the fact that the RG evolution equation at large $\theta$ or
large $N$ is of first order in the derivatives of the
generating functional with respect to the sources.

The non-linear parametric equation (\ref{parametric2}) implies
the following linear equation for the generating functional
of the planar amplitudes $\Gammapis_\Lambda^{({\rm pl})}$, the part
of $\Gammapis_\Lambda$ linear in the sources $\Omega_k$:
\be
\bigl[\partial_{\sigma_0} - 
2\,\partial_{\sss m_0^2}\Gammapis_\Lambda^{\sss \{0\}}\partial_{\sss m_0^2}\bigr]\,\Gammapis_\Lambda^{({\rm pl})}=0
\label{linearparametric}
\ee
where $\Gammapis_\Lambda^{\sss \{0\}}$ is the vacuum energy density,
the part of $\Gammapis_\Lambda$ independent of the sources. 

We want now to translate the ``bare''  equation (\ref{linearparametric})
into a renormalized equation.
Let us start first with the massive theory. The renormalized
generating functional $\Gammapis_{\rm ren}\bigl[\rho^{\sss (a)};\Lambda,
\Lambda_R\bigr]$ depends on the renormalized couplings $\rho^{\sss (a)}$
both through the bare ones and also, as far as the mass is concerned, 
explicitly via the propagators: 
\be
\Gammapis_{\rm ren}\bigl[\rho^{\sss (a)};\Lambda, \Lambda_R\bigr]
\equiv \Gammapis_{\Lambda,\,\Lambda_0}\bigl[
\rho_0^{\sss (a)}\bigl(\rho^{\sss (a)}; \Lambda_R, \Lambda_0\bigr); m^2\bigr]
\label{renormalized}
\ee
Thus
\be
\partial_{\sss \rho_0^{\sss (a)}} =
{\partial\rho^{\sss (b)}\over \partial\rho_0^{\sss (a)} }\,
\Bigl[\partial_{\rho^{\sss (b)}} -\delta_{b,{\sss m^2}}
\,\partial_{\sss m^2}\Bigr]
\ee
where $\partial_{\sss m^2}$ is the derivative with respect to
the explicit $m^2$ dependence of the generating functional. Hence 
\be
\Bigl[\bigl(\partial_{\sigma_0} - 
2\,\partial_{\sss m_0^2}\Gammapis_\Lambda^{\sss \{0\}}\partial_{\sss m_0^2}\bigr)\,\rho^{\sss (b)}\Bigr]\,
\Bigl[\partial_{\rho^{\sss (b)}} -\delta_{b,{\sss m^2}}\,\partial_{\sss m^2}
\Bigr]\Gammapis_{\rm ren}^{\rm pl}\bigl[\rho^{\sss (a)};\Lambda, \Lambda_R \bigr]=0
\label{massiveparametricmess}
\ee
The previous equation  simplifies at the renormalization scale
$\Lambda =\Lambda_R$. Indeed, acting on Eq. (\ref{linearparametric}) with the 
normalization operators ${\rm N}_b$ (with
$b\in \{m^2, Z, g,\sigma\}$), one obtains
\be
\bigl(\partial_{\sigma_0} - 
2\,\partial_{\sss m_0^2}\Gammapis_{\Lambda_R}^{\sss \{0\}}
\partial_{\sss m_0^2}\bigr)\,{\rm N}_b[\Gammapis_{\Lambda_R}^{\rm pl}]=
\bigl(\partial_{\sigma_0} - 
2\,\partial_{\sss m_0^2}\Gammapis_{\Lambda_R}^{\sss \{0\}}
\partial_{\sss m_0^2}\bigr)\,\rho^{\sss (b)}= 0
\ee
for all planar couplings $\rho^{\sss (b)}$, i.e. for $b\not=\sigma$. 
Thus 
Eq. (\ref{massiveparametricmess}) reduces for $\Lambda=\Lambda_R$ 
to
\be
\partial_\sigma\,\Gammapis_{\rm ren}^{\rm pl}
\bigl[\rho^{\sss (a)};\Lambda_R, \Lambda_R\bigr]=0
\label{parametricmassive}
\ee
since $\bigl(\partial_{\sigma_0} - 
2\,\partial_{\sss m_0^2}\Gammapis_{\Lambda_R}^{\sss \{0\}}
\partial_{\sss m_0^2}\bigr)\,\sigma \not=0$.
The parametric equation (\ref{parametricmassive}) shows that 
planar amplitudes of the massive theory
evaluated at the renormalization scale $\Lambda_R$ are
independent of the non-planar coupling $\sigma$. For example,
from Eq. (\ref{2pointsrenormalized}), one sees that 
$\partial_\sigma\Gamma^{\sss \{2\}}_{\Lambda,\Lambda_R}$
vanishes at $\Lambda=\Lambda_R$. The parametric equation 
(\ref{parametricmassive}) also means that if we consider
$\Gammapis_{\rm ren}^{\rm pl}$ at $\Lambda=\Lambda_R$
as function of the renormalized $m^2, Z, g$ and of
the {\it bare} $\sigma_0$, it does not depend on $\sigma_0$.
In other words in the massive theory
one can forget about the $\sigma$ coupling
if one only wants to compute planar amplitudes at $\Lambda=\Lambda_R$.

The massless case is more subtle and thus more interesting.
Because of the massless renormalization condition (\ref{renconditionsmassless})
there is one less renormalized coupling than there are bare couplings.
The renormalization conditions for $Z$, $g$ and $\sigma$
in Eq. (\ref{renormalizationconditions}) 
express the renormalized couplings $\rho^{\sss(\a)}$, where
$\a \in \{Z,g,\sigma\}$, as functions of the bare couplings 
$\rho^{\sss (\b)}_0$ and $m_0^2$:
\be
\rho^{\sss (\a)}= \rho^{\sss (\a)}(\rho^{\sss (\b)}_0, m_0^2;
\Lambda_R, \Lambda_0)
\label{renversusbare}
\ee
where $\Lambda_R > 0$ to regulate the infrared divergences.
The massless renormalization condition (\ref{renconditionsmassless})
determines $m_0^2 = m_0^2(\rho^{\sss (\b)}_0;\Lambda_0)$ as function of
the bare $\rho^{\sss (\b)}_0$: by substituting this latter expression
into Eq.  (\ref{renversusbare}) one obtains the renormalized 
$\rho^{\sss (\a)}$ as functions of the bare  $\rho_0^{\sss (\b)}$:
\be
\rho^{\sss (\a)}=
\rho^{\sss (\a)}(\rho^{\sss (\b)}_0, m_0^2(\rho^{\sss (\b)}_0;\Lambda_0);
\Lambda_R,\Lambda_0)
\label{renversusbare2}
\ee
From now on when writing $\rho^{\sss (\a)}$ 
we refer to the
functions of $\rho^{\sss (\b)}_0$ defined in the equation above:
denoting by $\rho_0^{(\b)}(\rho^{(\a)};\Lambda_R,\Lambda_0)$ their inverses, 
the renormalized functional is defined by
\be
\Gammapis_{\rm ren}\bigl[\rho^{\sss (\a)};\Lambda, \Lambda_R\bigr]
\equiv\Gammapis_{\Lambda,\,\Lambda_0}\bigl[
\rho_0^{\sss (\b)}\bigl(\rho^{\sss (\a)}; \Lambda_R, \Lambda_0\bigr),
m_0^2(\rho^{\sss (\b)}_0(\rho^{\sss (\a)};\Lambda_R,\Lambda_0);\Lambda_0)\bigr]
\label{renormalizedmassless}
\ee
Therefore
\be
\partial_{\rho^{\sss (\a)}}= {\partial\rho_0^{\sss (\b)}\over
\partial\rho^{\sss (\a)}}\,{\partial\over\partial\rho_0^{\sss (\b)}} + 
{\partial m_0^2\over
\partial\rho^{\sss (\a)}}\,{\partial\over\partial m_0^2}
\label{bareversusren}
\ee
and thus 
\be
{\partial\rho^{\sss (\a)}\over\partial \sigma_0}
{\partial\Gammapis_{\rm ren}^{\rm pl}\over \partial\rho^{\sss (\a)}}= {\partial
\Gammapis^{\rm pl}_\Lambda\over\partial\sigma_0} + 
{\partial\rho^{\sss (\a)}\over\partial \sigma_0}\,{\partial m_0^2\over
\partial\rho^{\sss (\a)}}\,{\partial\Gammapis^{\rm pl}_\Lambda
\over\partial m_0^2}= 
{\partial\Gammapis^{\rm pl}_\Lambda\over\partial\sigma_0} + 
{\partial m_0^2\over\partial \sigma_0}\,{\partial\Gammapis^{\rm pl}_\Lambda
\over\partial m_0^2}
\ee
Using the parametric bare equation (\ref{linearparametric})
we obtain
\be
{\partial\rho^{\sss (\a)}\over\partial \sigma_0}
{\partial\Gammapis_{\rm ren}^{\rm pl}
\over \partial\rho^{\sss (\a)}}[\Lambda,\Lambda_R]=
\Bigl[2\,\partial_{\sss m_0^2}\Gammapis_\Lambda^{\sss \{0\}} + 
{\partial m_0^2\over\partial \sigma_0}\Bigr]\,
{\partial\Gammapis_\Lambda^{\rm pl}\over\partial m_0^2}
\label{parametricmassless2}
\ee
To evaluate ${\partial m_0^2\over\partial \sigma_0}$ we make 
use of the massless renormalization condition (\ref{renconditionsmassless}),
which gives
\be
{\partial m_0^2\over\partial \sigma_0}= -\Biggl[{\partial\, {\rm N}_{\sss m^2}
[\Gammapis_{\sss \Lambda =0}]\over \partial m_0^2}\Biggr]^{-1}\, 
{\partial\, {\rm N}_{\sss m^2}
[\Gammapis_{\sss \Lambda =0}]\over \partial\sigma_0}
\ee
Substituting into Eq. (\ref{parametricmassless2}), one gets
\be
{\partial\rho^{\sss (\a)}\over\partial \sigma_0}
{\partial\Gammapis_{\rm ren}^{\rm pl}
\over \partial\rho^{\sss (\a)}}[\Lambda,\Lambda_R]= 
-{\partial \Gammapis_\Lambda^{\rm pl}\over\partial m_0^2}\,
\Biggl[{\partial\, {\rm N}_{\sss m^2}
[\Gammapis_{\sss \Lambda =0}]\over \partial m_0^2}\Biggr]^{\scriptstyle -1}
\!\Bigl[\partial_{\sigma_0}- 2\,\partial_{\sss m_0^2}\Gammapis_\Lambda^{\sss \{0\}}\,\partial_{\sss m_0^2}\Bigr]\,
{\rm N}_{\sss m^2}[\Gammapis_{\sss \Lambda =0}]
\label{parametricmassless3}
\ee
This equation simplifies considerably
when evaluated at $\Lambda =0$, since then the R.H.S.
becomes proportional to the bare parametric equation and hence vanishes:
\be
R^{\sss(\a)}\,{\partial\Gammapis_{\rm ren}^{\rm pl}
\over \partial\rho^{\sss (\a)}}[\Lambda,\Lambda_R]\big|_{\Lambda=0}
= 0
\label{parametricmassless4}
\ee
where   
\be
R^{\sss (\a)} \equiv {\partial\rho^{\sss (\a)}\over\partial \sigma_0}
\label{betafunctions}
\ee
The derivative of 
$\Gammapis_{\rm ren}^{\rm pl}$ with respect to $Z$ which appear
in Eq. (\ref{parametricmassless4}) can be replaced by $g$ and $\sigma$
derivatives by using the so-called counting identity. 
This identity takes the following simple form when evaluated at $\Lambda=0$
\be
Z\partial_Z\,\Gammapis^{\sss \{k\}}_{\rm ren}[0,\Lambda_R] =
\Bigl[{k\over 2} - 2 (g\partial_g + \sigma\partial_\sigma)\Bigr]
\,\Gammapis^{\sss \{k\}}_{\rm ren}[0,\Lambda_R]
\label{countingidentity}
\ee
where $\Gammapis^{\sss \{k\}}_{\rm ren}[\Lambda,\Lambda_R]$ is the
planar amplitude with $k$ external legs.
Therefore, choosing $Z=1$, one can rewrite Eq.  (\ref{parametricmassless4}) 
as follows
\be
\bigl[\partial_\sigma +\chi\,\partial_g + \tau \bigr]\,
\Gammapis_{\rm ren}^{\sss \{k\}}[0,\Lambda_R]= 0
\label{parametricmassless5}
\ee
where we introduced the generalized beta-functions
\be
\chi(g,\sigma) = {R^{\sss (g)} - 2\, g\, R^{\sss (Z)}\over
R^{\sss (\sigma)} - 2\, \sigma\, R^{\sss (Z)}}\qquad\quad
\tau (g,\sigma) = {1\over 2}\, {k\, R^{\sss (Z)}\over
R^{\sss (\sigma)} - 2\, \sigma\, R^{\sss (Z)}}
\label{chifunctions}
\ee
We will see in a moment that  $\chi$ and $\tau$ get their
first non-vanishing contributions at 2 and 3 loops respectively. 
The generalized beta-functions (\ref{chifunctions}) capture therefore 
the ``anomalous'' dependence
of the planar amplitudes on the non-planar renormalized coupling $\sigma$
in the massless theory: this effect, as it will be apparent
from the computation that follows,  is due to an interplay between
the UV and IR divergences of the theory.

\begin{figure}
\begin{center}
\includegraphics*[scale=.3, clip=false]{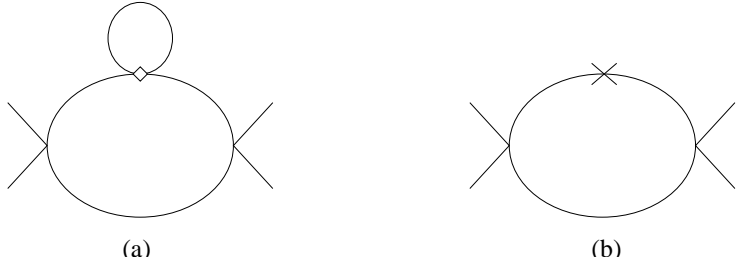}
\end{center}
\caption[x] {\footnotesize $\sigma$-dependent 2-loops corrections to 
$\Gamma^{\sss\{4\}}$. The diamond is the $\sigma$ vertex and the cross
the mass counterterm insertion}
\label{f:diagrams4}
\end{figure}

From Eqs. (\ref{2loopsbarecouplings}) it follows that
\be
R^{\sss (Z)} = O(\hbar^3)\qquad\qquad R^{\sss (\sigma)} = 1 + O(\hbar)
\ee
The first non-vanishing contributions to $R^{\sss (g)}$ 
come from the 2 loop
diagrams in Figure~\ref{f:diagrams4}. Therefore $\tau (g,\sigma) = O(\hbar^3)$
and 
\be
\chi (g, \sigma) = R^{\sss (g)}_{\sss 2\, loops} +O(\hbar^3)
\ee
Up to order $O(\hbar^3)$ 
the $\sigma_0$ dependence of $g$ is given by the expression
\be
g = g_0 + 2\, g_0^2 \, I^{(2)}_{\Lambda_R,\, \Lambda_0}(0) +
4\, \sigma_0\,g_0^2\, I^{(1)}_{\Lambda_R,\,\Lambda_0}\, 
I^{(4)}_{\Lambda_R,\, \Lambda_0}(0) +4\, m^2_0\,g_0^2\,
I^{(4)}_{\Lambda_R,\, \Lambda_0}(0) + O(g_0^3)
\label{g2loops}
\ee
where
\be
I^{(4)}_{\Lambda_R,\, \Lambda_0}(p)\equiv \int\!{d k\over (2\pi)^4}\,
(\Delta_{\Lambda,\,\Lambda_0}(k))^2\,\Delta_{\Lambda,\,\Lambda_0}(k-p)
\ee
$m^2_0$ as a function of $\sigma_0$
and $g_0$ at 1 loops is determined by the first of 
Eqs. (\ref{2loopsbarecouplings}) {\it evaluated for $\Lambda_R=0$}:
\be
m^2_0 = -(\sigma_0 +2\, g_0) I^{(1)}_{0,\,\Lambda_0} +O(\hbar^2)
\label{masslessness}
\ee
Plugging this expression into  Eq. (\ref{g2loops}) one obtains for the
beta function $\chi(g,\sigma)$ at two loops the following result
\be
R^{\sss (g)}_{\sss 2\, loops} = -4\,g^2\, I^{(1)}_{0,\,\Lambda_R}\, 
I^{(4)}_{\Lambda_R,\, \Lambda_0}(0)= - {4\,g^2\over (16\pi^2)^2}\,\Lambda_R^2\,
\Bigl({1\over \Lambda_R^2}-{1\over \Lambda_0^2}\Bigr)\,
{\mathop{\longrightarrow}^{\sss \Lambda_0\to\infty} }\, -{g^2\over 64\,\pi^4}
\ee
Let us verify the massless parametric equation (\ref{parametricmassless5})
for the planar 4-point function  at 2 loops, the lowest order for
which the equation is non-trivial. Let us choose, just for simplicity,
the external momenta $(p_1,p_2,p_3,p_4)$ equal to $(p,0,-p,0)$. 
The bare 4-point function becomes:
\bea
&&\!\!\!\!\!\!\!\!\!\!\!\!\!\!\!\!\!\!
\Gamma^{\sss \{4\}}_{\Lambda,\,\Lambda_0}(p) = g_0 + 2\,g_0^2\,
I^{(2)}_{\Lambda,\, \Lambda_0}(p) +
4\,g_0^2\,\Bigl(\sigma_0\, I^{(1)}_{\Lambda,\,\Lambda_0} +
 m^2_0\Bigr)\,I^{(4)}_{\Lambda,\, \Lambda_0}(p) + O(g_0^3) = \nonumber\\
&&\quad\! =  g_0 + 2\,g_0^2\,
I^{(2)}_{\Lambda,\, \Lambda_0}(p) -
4\, \sigma_0\,g_0^2\,I^{(1)}_{0,\,\Lambda}\,
I^{(4)}_{\Lambda,\, \Lambda_0}(p) + O(g_0^3)
\eea
where in the second line of the equation above we used the masslessness 
constraint (\ref{masslessness}). Substituting now bare with renormalized 
couplings (using Eq. (\ref{g2loops})), we find that the renormalized 4-point 
function is given by the $\Lambda_0\to\infty$ limit of the following 
expression:
\bea
&&\!\!\!\!\!\!\!\!\!\!\Gamma^{\sss \{4\}}_{\Lambda,\,\Lambda_0;\,\Lambda_R}(p) = g - 2\,g^2\,\Bigl( I^{(2)}_{\Lambda,\, \Lambda_0}(p)-
I^{(2)}_{\Lambda_R,\, \Lambda_0}(0)\Bigr) + \nonumber\\
&&\qquad\qquad-
4\,\sigma\,g^2\,\Bigl(I^{(1)}_{0,\,\Lambda}\,I^{(4)}_{\Lambda,\, \Lambda_0}(p)-
I^{(1)}_{0,\,\Lambda_R}\,
I^{(4)}_{\Lambda_R,\, \Lambda_0}(0)\Bigr) + O(g^3)
\eea
We thus see that the L.H.S. of the parametric equation 
(\ref{parametricmassless5}), applied to the 4-point planar function,
equals
\be
\bigl[\partial_\sigma +\chi\,\partial_g + \tau \bigr]\,
\Gamma^{\sss \{4\}}_{\rm ren}(p)[0,\Lambda_R] = 
-4\,g^2\,\lim_{\Lambda\to 0\atop }\,
I^{(1)}_{0,\,\Lambda}\,I^{(4)}_{\Lambda,\, \infty}(p)
\label{parametriccheck}
\ee
Since
\be
-4\, g^2\,I^{(1)}_{0,\,\Lambda}\,I^{(4)}_{\Lambda,\, \infty}(p) =
{g^2\over (16\,\pi^2)^2}\,{\Lambda^2\over p^2}\,\Bigl[\log\,
\Bigl({\Lambda^2\over p^2}\Bigr)-1 + O\Bigl({\Lambda^2\over p^2}\Bigr)\Bigr] 
\ee
the R.H.S. of the Eq. (\ref{parametriccheck}) vanishes. Note that the
$\Lambda\to 0$ limit of Eq. (\ref{parametriccheck}) must be defined by
taking $\Lambda/ p\to 0$.
\section{Conclusions}

The main message of this article is that, even at $\theta=\infty$,
renormalized non-commutative field theories do not reduce simply to
their planar sector. The genus zero RG equation couples planar amplitudes
to partially integrated non-planar amplitudes. Since the dependence of
PIS amplitudes on $\theta$ factors out, these amplitudes are
essentially the same as in matrix field theory.  While the non-planar
coupling $\sigma$ can be introduced in a local way in $N\times N$ matrix
field theory at finite $N$  --- via the multi-trace operator $\int dx ({\rm Tr}
\phi^2 (x))^2$ --- this is not possible in the non-commutative theory
of the Moyal type, since in this kind of theories every trace must be
accompanied by integration over non-commutative space.  The
Wilson-Polchinski genus zero equation allows for a perfectly rigorous
treatment of the non-local, non-planar ``bare'' coupling in Eq.
(\ref{sigmacounterterm}) and elucidates the mechanism by which this
non-local counterterm cancels local (planar) divergences of  planar
amplitudes.

The genus zero RG equation also clarifies the standing and the
limitations of the purely planar renormalization~\cite{seiberg} of
non-commutative field theories.  A distinguishing feature of the genus
zero RG equation is of being of first order in source derivatives: we
showed that this entails factorization, a well-known property of large
$N$ matrix models. One might expect, because of
factorization, to be able to disregard the non-planar
counterterms altogether when computing planar amplitudes
in the limit $\theta\to\infty$ 
(or, in the matrix field theory case, in the limit $N\to\infty$). 
We showed that this is not quite so. In the massive
theory factorization implies that one can indeed remove UV divergences
of planar amplitudes using only planar counterterms, if one keeps the
floating Polchinski cut-off $\Lambda$ equal to the renormalization
scale $\Lambda_R$ that defines the renormalized couplings.  In the
usual, commutative, situation UV finiteness of
$\Gamma_{\Lambda,\Lambda_R}$ at $\Lambda=\Lambda_R$ would imply its
finiteness for any $\Lambda$, since the Wilson-Polchinski RG equation
is essentially independent of the UV cut-off. In the non-commutative
case instead this is not the case: if one insists on introducing only
planar counterterms as soon as $\Lambda$ differs from $\Lambda_R$, the
UV scale $\Lambda_0$ reappears in the effective planar action. The
reason, of course, is that the RG equation does not close on planar
amplitudes and the $\Lambda$ derivative of a planar amplitudes
involves non-planar diagrams evaluated at exceptional momenta (see
Figure~\ref{f:diagrams2}), which --- in planar renormalization ---
are divergent. The situation for the massless theory is even more dramatic: 
planar counterterms are not enough to eliminate UV divergences even
if one sends the floating cut-off to the renormalization scale,
i.e. even in the limit $\Lambda\to 0$. 

The restriction to partially integrated amplitudes also elucidates
the nature of non-planar contributions to non-commutative current algebra
anomalies \cite{theisen}. The non-planar part of the topological charge 
which captures the axial non-commutative anomaly appears in
the effective action as a partially integrated non-local 
term of the type in Eq. (\ref{sigmacounterterm}). 
To give an example in 2 dimensions, let  $A_\mu$ be
the axial vector field to which the axial current is coupled, $V_\mu$
the background vector field which couples to the vector
current and $F_{\mu\nu}$ the field strength relative to $V_\mu$. 
Then the non-planar anomaly computed in
in \cite{theisen} is reproduced by a 
term in the effective action that, written in (non-commutative) configuration
space, writes as
\be
\lim_{\Lambda\to 0} \Lambda^2 \int\!\! d^2x\, {{\rm e}^{-\Lambda^2\, x^2}
\over 2\pi^2} {1\over {\partial}^{\,2}}\, \partial_\mu A_\mu(x) \int\!d^2y\, 
\epsilon^{\mu\nu} F_{\mu\nu}(y)
\ee
very much analogous to the partially integrated non-local term
in Eq. (\ref{sigmacounterterm}).

One can think of several possible extensions of our work. 
The most challenging is the construction of a renormalized
theory of partially integrated amplitudes at higher
genus. The analysis of partially
integrated amplitudes of genus $g$, that we present in Appendix B,
shows that these amplitudes go as $1/{\theta}^{d g}$ for large
$\theta$ --- unlike amplitudes at generic external momenta which do
not have a $\theta=\infty$ limit uniform in the external momenta.
This strongly suggests that renormalized partially integrated amplitudes 
at higher genus admit a meaningful $1/{\theta}$ 
expansion. Attacking the renormalization problem of higher genus amplitudes
necessitates first of all working out the corresponding expansion of the 
Wilson-Polchinski RG equation. The RG evolution of higher genus partially 
integrated amplitudes involves lower genus amplitudes with non-vanishing
total momenta flowing into two of their holes: for this reason
it seems that the understanding of the higher genus non-commutative theory 
might require significantly extending the ideas presented in this paper.
It is a problem that we leave for the future.
Another issue which emerges from the present
work is the interpretation of the restriction to partially 
integrated amplitudes from string theory point of view. One might 
also consider extending our analysis to gauge non-commutative theories.

\section*{Acknowledgments} 
We are glad to thank Prof. A. Schwimmer for in-depth discussions of
several aspects of this work.  This work is supported in part by
Ministero dell'Universit\`a e della Ricerca Scientifica e Tecnologica
and the European Commission's Human Potential program under contract
HPRN-CT-2000-00131 Quantum Space-Time, to which the authors are
associated through the Frascati National Laboratory.

\appendix
\section{Moyal Phases for Spherical Amplitudes}

In this appendix we derive a formula for the $\theta$ dependence
of generic spherical Feynman diagram integrands. 

Recall that planar diagrams depend on $\theta$ via the Moyal factor
\be
{\rm e}^{-i\, \Phi_n(p_1,\ldots, p_n)} \equiv {\rm e}^{-i\sum_{i<j} p_i\wedge p_j}
\label{moyalfactor1}
\ee
where $p_1\wedge p_2 \equiv {1\over 2}\theta_{\mu\nu}\,p_1^\mu\, p_2^\nu$, 
and $p_1,\ldots,p_n$ are the momenta associated to the $n$ external
lines of the graph. Let us briefly review the derivation of 
Eq. (\ref{moyalfactor1}) which 
exploits the following property
of planar double-line graphs: the momentum through any
propagator (or external line) in the graph can be written as
the difference $l_i -l_j$ 
where $l_i$ and $l_j$ are pseudo-momenta associated with the (oriented)
single lines that are the adjacent edges of the double-line propagator.
For any vertex with $k$ legs, let the momenta entering the vertex
be $q_1, q_2, \ldots, q_k,$ in cyclic
order: with respect to the commutative
theory, the Feynman rules of the Moyal non-commutative theory 
include the additional phase factor
\be
{\rm e}^{-i\, \Phi_k(q_1,\ldots, q_k)} = {\rm e}^{-i\sum_{i<j} q_i\wedge q_j}
\label{moyalfactor}
\ee
Writing the momenta $q_j$ in terms of pseudo-momenta, 
$q_j = l_{i_j}- l_{i_{j+1}}$, one obtains
\be
\sum_{i<j} q_i \wedge q_j =
l_{i_1}\wedge l_{i_2} +l_{i_2} \wedge l_{i_3} + \cdots + l_{i_n}
\wedge l_{i_1}
\ee
Thus the phase factor at any interaction point may be expressed 
as the product of $k$ terms, one for each incoming propagator
\be
{\rm e}^{-i\, \Phi_k(q_1,\ldots, q_k)}= 
\prod_{j=1}^{k}e^{-i\,(l_{i_j} \wedge\, l_{i_{j+1}})}
\ee
Any internal propagator gives two contributions  to the total phase factor
(\ref{moyalfactor1}) ---  one for each of its two end vertices ---  which
cancel each other.  Therefore only the external momenta contribute 
to the total phase factor, and one obtains Eq. (\ref{moyalfactor1}).

The representation of propagator momenta in
terms of pseudo-momenta is valid not only for the planar diagrams but,
more generally, also for PIS amplitudes. Therefore the very same 
argument which leads to Eq. (\ref{moyalfactor1}) 
generalizes to PIS amplitudes with $h$ holes: 
every hole with $k$ external lines gives a
phase factor ${\rm e}^{-i\Phi_k}$, and hence the total Moyal phase of
the amplitude is the product of $h$ factors like in
(\ref{moyalfactor1}), one for each hole. 

Let us now turn to generic non-planar spherical diagrams with 
$h$ holes. Let $a=1,\ldots, h$ be a label defining 
an arbitrary order of the holes.
Let $p^{(a)}_{i_a}$, with $i_a=1,\ldots, n_a$, be the 
$i_a$-th momentum entering the $a$-th hole of the spherical amplitude.
The $n_a$ momenta entering the hole have a natural cyclic order determined
by the orientation of the associated Riemann surface:
thus, defining the indices $i_a$ requires choosing a particular 
(first) momentum for each hole. A spherical diagram with $L$ loops
defines a triangulation of the sphere with $F= L+1$ faces. We called
holes the faces
to which external lines are attached, and thus, obviously, $h\le F=L+1$.
We denote by $q_A$, with $A=1,\ldots, L$ the independent loop momenta, 
arbitrarily chosen. The Moyal phase  $\Phi\, (q_A, p^{(a)}_{i_a})$
of the diagram has the general structure
\be
\Phi\,(q_A, p^{(a)}_{i_a}) =\! \sum_{A,B}C_{AB}\, q_A\wedge\,q_B +
\!\sum_{a,\,A,\, i_a}\!\! D^{(a)}_{A\,i_a} q_A \wedge p^{(a)}_{i_a} + 
\!\sum_{a,\, b,\, i_a,\, i_b}\!\!\!\!\!
E^{(a b)}_{i_a i_b}\, p^{(a)}_{i_a}\wedge\, p^{(b)}_{i_b}
\label{generalmoyal}
\ee
where $C_{AB}$, $D^{(a)}_{A\,i_a}$ and 
$E^{(a b)}_{i_a i_b}$ are constant coefficients.

When all the external momenta $p^{(a)}_{i_a}$ vanish, the amplitude
becomes planar with no external lines and, thus, the Moyal phase
vanish: it follows that $C_{AB}=0$. Also, if all the loop
momenta $q_A$ vanish, the resulting Moyal phase is
that of the  tree ---
and hence planar --- diagram obtained from the original diagram by 
cutting all the internal propagators associated with the momenta
$q_A$. This tree diagram has the external lines of the
original diagram, with an ordering which depends on the choice
of the cut propagators, i.e. on the choice of the $L$ independent
loop momenta $q_A$. Our choice of ordering of the holes and of
the ``first'' momenta of each hole (implicit in the definition of
the indices $a$ and $i_a$) induces, of course, an ordering on
the external momenta: we can always take this ordering to coincide
with the ordering of the external momenta of the tree
diagram above. In other words, the (arbitrary) definition of
the loop momenta $q_A$ should be consistent with the
(arbitrary) definition of the indices $a$ and $i_a$. With this
understanding, $\Phi(0, p^{(a)}_{i_a})$ writes as
\be
\Phi(0, p^{(a)}_{i_a})=\!\sum_{a,\, b,\, i_a,\, i_b}\!\!\!\!\!
E^{(a b)}_{i_a i_b}\, p^{(a)}_{i_a}\wedge\, p^{(b)}_{i_b} =
\sum_{a}\!\sum_{i_a<j_a}\! p^{(a)}_{i_a}\wedge\, p^{(a)}_{j_a}
+ \sum_{a<b}\Sigma^{(a)}\wedge\,\Sigma^{(b )}
\ee
where $\Sigma^{(a)}\equiv \sum_{i_a}\!p^{(a)}_{i_a}$ is the momentum
entering the $a$-th hole. Last, let us take all the $\Sigma^{(a)} =0$:
the amplitude becomes PIS and thus its Moyal phase reduces to 
$\sum_{a}\!\sum_{i_a<j_a}\! p^{(a)}_{i_a}\wedge\, p^{(a)}_{j_a}$.
Hence the term $\sum_{a,\,A,\, i_a}\!\! D^{(a)}_{A\,i_a} q_A \wedge p^{(a)}_{i_a}$ in (\ref{generalmoyal}) must vanish when $\Sigma^{(a)} =0$ and
therefore it  can be expressed as linear combination of the $\Sigma^{(a)}$ 
\be
\sum_{a,\,A,\, i_a}\!\! D^{(a)}_{A\,i_a} q_A \wedge p^{(a)}_{i_a}=
\!\sum_a k^{(a)} \wedge\, \Sigma^{(a )}
\ee
where $k^{(a)}$ are $h$ linear combinations of the $L$ loop momenta $q_A$.
In conclusion the Moyal phase of the diagram is
\be
\Phi\,(k^{(a)}, p^{(a)}_{i_a}) = 
\sum_{a}\!\sum_{i_a<j_a}\! p^{(a)}_{i_a}\wedge\, p^{(a)}_{j_a}
+ \sum_{a<b}\Sigma^{(a)}\wedge\,\Sigma^{(b )}+
\!\sum_a k^{(a)} \wedge\, \Sigma^{(a )}
\label{nonplanarmoyal}
\ee
The last term in the R.H.S. of Eq. (\ref{nonplanarmoyal})
gives an IR sensitive UV cut-off for the integrand of the
corresponding Feynman amplitude --- the origin
of the famous IR-UV mixing effect. Because of this term, the Feynman
integral is not --- for $\Sigma^{(a)}\not=0$ --- an analytic function
of the non-commutative parameter $\theta$ at $\theta=\infty$.  
As we mentioned above, PIS
amplitudes  (for which $\Sigma^{(a)}=0$) are precisely those that 
admit a good $\theta\to\infty$ limit: for them, 
the R.H.S. of Eq. (\ref{nonplanarmoyal}) reduces to the first term,
which, thanks to momentum conservation, is now independent of
the ordering choices underlying the definition of the indices
$a$ and $i_a$. 

\begin{figure}
\centerline{
\begin{picture}(100,100)(0,0)
\put(-1,15){\makebox(0,0){\tiny
3}}\put(42,49){\makebox(0,0){\tiny 4}} \put(76,15){\makebox(0,0)
{\tiny 4}}\put(-1,92){\makebox(0,0){\tiny 2}}
\put(76,92){\makebox(0,0){\tiny 1}} \put(31,49){\makebox(0,0){\tiny 1}}
\put(31,59){\makebox(0,0)
{\tiny 2}}\put(42,59){\makebox(0,0){\tiny 3}}
\put(7,29){\line(0,1){49}}\put(7,29){\line(-1,-1){10}}\put(7,78)
{\line(-1,1){10}}\multiput(7,25)(-2,-2){4}{.}\multiput(8,80)(-2,2){4}{.}
\put(12,24){\line(1,0){49}}\put(12,24){\line(-1,-1){10}}\put(12,84)
{\line(1,0){49}}\put(12,84){\line(-1,1){10}}
\put(17,39){\line(1,1){10}}\put(17,39){\line(0,1){30}}\put(17,69)
{\line(1,-1){10}}\multiput(16,35)(2,2){6}{.}\multiput(17,72)(2,-2){6}{.}
\put(23,34){\line(1,0){29}}\put(23,34){\line(1,1){10}}\put(23,74)
{\line(1,0){29}}\put(23,74){\line(1,-1){10}}
\put(52,34){\line(-1,1){10}}\put(52,74){\line(-1,-1){10}}
\put(57,39){\line(0,1){30}}\put(57,39){\line(-1,1){10}}\put(57,70)
{\line(-1,-1){10}}\multiput(54,72)(-2,-2){6}{.}
\put(62,24){\line(1,-1){10}}\put(61,84){\line(1,1){10}}
\multiput(64,82)(2,2){4}{.}
\put(67,29){\line(0,1){49}}\put(67,29){\line(1,-1){10}}
\put(67,78){\line(1,1){10}}
\multiput(10,26)(3,0){18}{.}\multiput(8,26)(0,3){18}{.}
\multiput(71,18)(-2,2){15}{.}\multiput(14,33)(0,3){15}{.}
\multiput(10,80)(3,0){18}{.}\multiput(15,75)(3,0){15}{.}
\multiput(57,33)(0,3){15}{.}
\end{picture}}
\caption[x] {\footnotesize A spherical non-planar 
diagram with two holes and its auxiliary momentum path}
\label{f:path}
\end{figure}
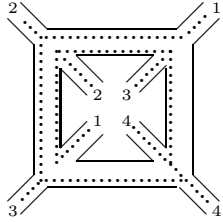

An explicit definition for the $h$ loop momenta $k^{(a)}$ in Eq. 
(\ref{nonplanarmoyal}) can be
given as follows. Take the original non-planar diagram and put to zero
all the external momenta $p^{(a)}_{i_a}$. The resulting diagram
is planar and its internal momenta admit the EK representation
in terms of pseudo-momenta $l_A$, with $A=1,\ldots L+1$, one
for each face of the diagram. To take into account the external
momenta consider also an auxiliary oriented 
path running through the double-line
propagators with the following properties: (i) the path
connects all the points to which the external legs are attached; (ii)
it turns clockwise around each hole starting from the arbitrary
chosen ``first'' leg to the ``last'' and going from the arbitrary chosen
``first'' hole to the ``last'' (thus defining an ordering of the
external legs); (iii) the path together with the external legs attached
to it forms a tree diagram whose propagators carry the momenta
which  enter through the external legs. An example of such
auxiliary momentum path is depicted in Figure~\ref{f:path}.

The EK prescription for
the momentum $k^{(a)}$ flowing through a given propagator is now
corrected by adding to the pseudo-momenta contribution the
momenta carried by the path, if this happens to go through
the propagator. With this definition of the internal $k^{(a)}$
the Moyal phase (\ref{nonplanarmoyal}) of a spherical non-planar
diagram writes as
\be
\Phi\,(l_a, p^{(a)}_{i_a}) = 
\sum_{a}\!\sum_{i_a<j_a}\! p^{(a)}_{i_a}\wedge\, p^{(a)}_{j_a}
+ \sum_{a<b}\Sigma^{(a)}\wedge\,\Sigma^{(b )}+
\!2\,\sum_a l_a \wedge\, \Sigma^{(a )}
\label{nonplanarmoyalek}
\ee
where $l_a$ with $a=1,\ldots, h\le L+1$ 
are the pseudo-momenta associated with the holes.
Since $\sum_a \Sigma^{(a)} =0$, $\Phi\,(l_a, p^{(a)}_{i_a})$
is invariant under $l_a\to l_a + c$, and thus depends only
on the differences of the pseudo-momenta $l_a$.

\section{Moyal phases for higher genus amplitudes}

In this appendix we will analyze the $\theta$ dependence of non-spherical
amplitudes. 

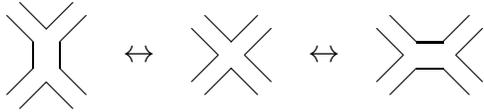
\begin{figure}
\centerline{
\begin{picture}(300,40)(-60,20)
\put(10,35){\line(0,1){10}}\put(10,35){\line(-1,-1){10}}
\put(10,45){\line(-1,1){10}}
\put(15,30){\line(-1,-1){10}}\put(15,30){\line(1,-1){10}}\put(15,50)
{\line(-1,1){10}}\put(15,50){\line(1,1){10}}
\put(20,35){\line(0,1){10}}\put(20,35){\line(1,-1){10}}\put(20,45)
{\line(1,1){10}}
\put(50,40){\makebox(0,0){$\leftrightarrow$}}
\put(80,40){\line(-1,-1){10}}\put(80,40){\line(-1,1){10}}
\put(85,35){\line(-1,-1){10}}\put(85,35){\line(1,-1){10}}\put(85,45)
{\line(-1,1){10}}\put(85,45){\line(1,1){10}}
\put(90,40){\line(1,-1){10}}\put(90,40){\line(1,1){10}}
\put(120,40){\makebox(0,0){$\leftrightarrow$}}
\put(150,40){\line(-1,-1){10}}\put(150,40){\line(-1,1){10}}
\put(155,35){\line(-1,-1){10}}\put(155,35){\line(1,0){10}}\put(155,45) 
{\line(-1,1){10}}\put(155,45){\line(1,0){10}}
\put(165,35){\line(1,-1){10}}\put(165,45){\line(1,1){10}}
\put(170,40){\line(1,-1){10}}\put(170,40){\line(1,1){10}}
\end{picture}
}
\caption[x] {\footnotesize Topological deformations preserving 
the Moyal phases}
\label{f:deformation}
\end{figure}

Let us begin with the following remark: 
the Moyal phase associated with a double-line diagram
is invariant under 
topological deformations of the type depicted in Figure~\ref{f:deformation}. 
These are deformations which vary the lengths of the double-line
propagators and correspond to changing the triangulation
of the underlying Riemann surface by keeping fixed its genus
$g$ and its number $F$ of faces. Note that the number of loops
$L$ of the Feynman diagram, which is given by
\be
L = F-1 +2\,g\, ,
\label{loopcount}
\ee
is left unchanged by these deformations.

\begin{figure}
\begin{center}
\includegraphics*[scale=.23, clip=false]{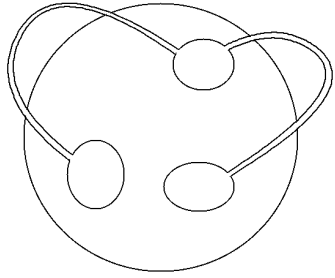}
\end{center}
\caption[x] {\footnotesize A genus 2 surface from a sphere with
3 holes and 4 external legs}
\label{f:highergenus}
\end{figure}

Consider now a
diagram of genus $g$ and $F$ faces and join with a double-line
propagator two external legs attached to two {\it different}
holes: one obtains in this way 
a diagram of genus $g+1$ and $F-1$ faces. This follows
from the Euler relation $2-2g= F-E+V$, where $V$, $E$ and $F$
are the numbers of vertices, propagators and faces of the diagram:
the new diagram has the same number of vertices, one more
propagator and one face less than the original diagram and,
thus, one more handle. Therefore we can build diagrams
of any genus $g$ and any number of faces $F$ starting from
spherical (non-planar) diagrams with $g+F$ faces by means
of the following construction: Consider
one such spherical diagram and join $g$ pairs of external legs
with $g$ propagators --- choosing the legs of each pair 
to belong to different holes. In other words there should be at
most a single double-line propagator connecting any pair of holes.
An example of this construction for a genus 2 surface
built out of a spherical diagram with 3 holes and 4 external
legs is given in Figure~\ref{f:highergenus}.

An important result in the theory of Riemann surfaces states
that double-line diagrams with fixed $g$ and $F$ 
provide a cell decomposition of the  moduli space of oriented 
Riemann surfaces of genus $g$ and $F$ boundaries: the moduli
of Riemann surfaces are parametrized by non-negative real numbers
associated with the lengths of the double-line propagators.
Since the moduli space of fixed genus and fixed number of boundaries
is a connected variety, it follows that
one can transform, by means of the deformations   
in Figure~\ref{f:deformation}, any graph of genus $g$ and $F$ faces
into a topologically equivalent one
built out of spherical non-planar diagrams in the way 
explained in the previous paragraph.

Given two topologically equivalent diagrams, their loop 
momenta are in a one-to-one correspondence and thus can be identified:
under this identification their Moyal phases coincide. We can therefore 
limit ourselves to evaluate the Moyal phase of the higher genus diagrams
built out of spherical non-planar graphs. The Moyal phases of such
graphs are given by the formula in Eq. (\ref{nonplanarmoyalek})
for the associated non-planar spherical graphs  
where some of the external momenta $p^{(a)}_{i_a}$  --- those flowing into
the legs which are joined by the $g$ propagators --- become loop momenta, 
$q_i$ (with $i=1,\ldots,g$) of the higher genus diagrams. Thus we can
write the sum $\Sigma^{(a)}$ of the momenta entering the $a$-th
hole, which appears in Eq. (\ref{nonplanarmoyalek}),
as follows:
\be
\Sigma^{(a)}= \sum_{i=1}^g q_i\, e_i^a + P^{(a)}
\label{external}
\ee 
In the formula above $P^{(a)}$ is momentum carried by the
external legs  of the higher genus diagram attached to the $a$-th hole
of the corresponding spherical diagram; $e_i^a$ is a 
numerical matrix whose $(i,a)$-th element
is +1 (-1) if the momentum $q_i$ enters (leaves) the $a$-th hole and
0 otherwise. $e_i^a$ is the incidence matrix of the
graph whose points
are the holes of the non-planar spherical diagram and whose lines
are the $g$ propagators which connect the holes. This is a
{\it tree} graph because any pair of holes is
connected at most by a single propagator.

The Moyal phase of the higher genus diagram is therefore a
quadratic form in the loop momenta $l_a$ and $q_i$ which
looks as follows:
\be
\Phi_{higher-genus} = \sum_{i,j} C_{ij}\, q_i \wedge\,q_j + 2 \sum_{i,\,a}
e_i^a\, l_a \wedge q_i +A
\ee
where $C_{ij}$ is a numerical matrix  and
$A$ is at most linear in the loop momenta. 
Since $\sum_{i,\,a} e_i^a\,
q_i =0$, the Moyal phase $\Phi_{higher-genus}$ depends only
on the differences of the $l_a$'s. In conclusion, the part of
$\Phi_{higher-genus}$ quadratic in the loop momenta $ Q_A \equiv (l_a, q_i)$ 
can be written as $\sum_{A,B} D_{AB}\, Q_A \wedge Q_B$ where 
$D_{AB}$ is an anti-symmetric matrix of the following form 
\be
(D)_{AB} = \left(\matrix{0 & e_j^b\cr -e_i^a & C_{ij}}\right)
\label{quadraticloop}
\ee
As we said above, $e_i^a$ is the incidence matrix of a tree 
graph with $g$ lines and thus it has rank $g$. It follows
that the matrix $D_{AB}$ has rank $2g$.

\begin{figure}
\begin{center}
\includegraphics*[scale=.23, clip=false]{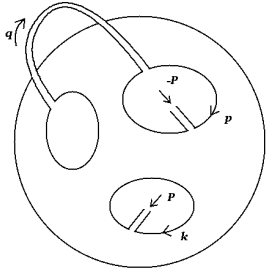}
\end{center}
\caption[x] {\footnotesize A genus 1 amplitude with
3 loops and 2 external legs}
\label{f:3loops}
\end{figure}

We are now ready to discuss the $\theta$ dependence of a
generic diagram of genus $g$. To understand the general
situation let us consider the example in Figure~\ref{f:3loops} of
a diagram of genus 1 and 3 loops
with 2 external legs carrying momentum $P$ and $-P$. 
By using the Schwinger parametrization for the
propagators one obtains a Feynman amplitude which writes
as follows:
\bea
&& \int\!\! d\a \, d\b \, 
d\c \, d\d \, d\eta\, d\zeta\, g(\a,\b,\c, \d, \eta, \zeta)\times\\
&&\quad\times\!\int\!d^dp\, d^dq\, d^dk\,
{\rm e}^{-\bigl[\a p^2+\b q^2+\c k^2+\d (p-q)^2+\eta
(p-k)^2+\zeta(q-k)^2 + 2 i p\wedge q + 2ik\wedge P\bigr]}\nonumber\\
&&\qquad \equiv \int\!\! d\a \, d\b \, 
d\c \, d\d \, d\eta\, d\zeta\, g(\a,\b,\c, \d, \eta, \zeta) \,
I(\a,\b,\c, \d, \eta, \zeta ;P)\nonumber
\label{3loopsamplitude}
\eea
Performing the integration over the loop momenta one obtains the following
function of the Schwinger parameters
\be
I(\a,\b,\c, \d, \eta, \zeta ;P) ={\pi^{3 d\over 2}\over 
\bigl(D+F\theta^2\bigr)^{d\over 2}}\,
{\rm e}^{-P^2\theta^2{E+\theta^2\over D+F\theta^2}}
\ee
where 
\bea 
&& D\equiv (\a+\b+\c)(\zeta\d+\zeta\eta+\d\eta)
+\a\b(\zeta+\eta)+\nonumber\\
&&\qquad\, + \b\c(\d+\eta)+\c\a(\d+\zeta)+\a\b\c\nonumber\\
&& E\equiv (\a+\eta)(\b+\zeta)+\d(\a+\b+\zeta+\eta)\nonumber\\ 
&& F\equiv \c+\eta+\zeta 
\label{schwingerfunctions}
\eea
If the function $g(\a,\b,\c, \d, \eta, \zeta)$ is sufficiently
regular at infinity, we can replace in Eq. (\ref{3loopsamplitude})
the integral $I$ with its asymptotic expression for $\theta\to\infty$ 
\be
I(\a,\b,\c, \d, \eta, \zeta ;P) \,\mathrel{\mathop\rightarrow_{\theta\to\infty}}\, {\pi^{3d\over 2}\over
F^{d\over 2}\theta^d}\,{\rm e}^{-{P^2\theta^2\over F}}
\ee
Note the $\theta$ dependence of this amplitude:
first of all there is a multiplicative factor $1/\theta^d$, which 
in the general case becomes $1/\theta^{dg}$. 
The non-trivial dependence on the Schwinger parameters of
the exponential factor ${\rm e}^{-{P^2\theta^2\over F}}$
is the source of the IR-UV mixing effect: 
if the external momenta are non-exceptional, $P\not=0$, the UV divergences
at $F=0$ are regulated by the UV cut-off $1/(P\theta)$.
This is what makes the limit $\theta\to\infty$ non-uniform in the
external momenta. 

\begin{figure}
\begin{center}
\includegraphics*[scale=.23, clip=false]{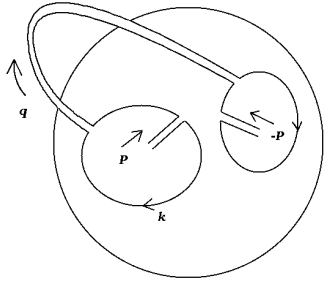}
\end{center}
\caption[x] {\footnotesize A genus 1 amplitude with
2 loops and 2 external legs}
\label{f:2loops}
\end{figure}

Note that when
the number $L$ of loop momenta equals $2g$ --- and thus the number
of faces of the higher genus diagram is 1 ---
the matrix $D_{AB}$ in Eq. (\ref{quadraticloop}) has {\it maximal} rank
and, thus, in this case the Moyal factor regulates all the loop integrations. 
For this special kind of diagrams the analogue of the function $F$ 
appearing in (\ref{schwingerfunctions}) does not vanish for any value
of the Schwinger parameters and hence there is no UV-IR mixing effect. 
For example consider the
amplitude, given in Figure~\ref{f:2loops}, 
of genus 1, 2 loops and 2 external legs carrying momentum $\pm P$:
\bea
&&\int\!\! d\a \, d\b \, 
d\c \, f(\a,\b,\c)\, \int\!d^dp\, d^dq
\,{\rm e}^{-\bigl[\a p^2+\b q^2+\c(p+q)^2+ 2ip\wedge q + 2ip\wedge P\bigr]}\nonumber\\
&&\qquad =\int\!\! d\a \, d\b \, 
d\c \, f(\a,\b,\c)\,{\pi^d\over\bigl[\a\b+\a\c+\b\c+\theta^2\bigr]^{d\over2}}\,{\rm e}^{-{\theta^2\,P^2(\beta +\gamma)\over \a\b+\a\c+\b\c+\theta^2}}\nonumber\\&&\qquad \mathrel{\mathop\rightarrow_{\theta\to\infty}}\,
{\pi^d\over\theta^d}\int\!\! d\a \, d\b \, 
d\c \, f(\a,\b,\c)\,{\rm e}^{-(\beta+\gamma)\,P^2}
\eea

Summarizing, in the $\theta\to\infty$  
limit, the integrated amplitudes go as
$\theta^{-dg}$, while the non-integrated ones
vanish exponentially. This is somewhat analogous
to what happens in $N\times N$ matrix field theories,
under the identification ${1\over N} \leftrightarrow {1\over \theta^d}$.  
In this analogy the contributions to the non-commutative
amplitudes coming from non-exceptional
external momenta correspond to
the ${\rm e}^{-N}$ non-perturbative instanton effects of matrix
theory.

\section{1PI RG equation in the planar limit}
In this appendix we derive the RG equation (\ref{1PIlargenrg}) satisfied
by the generating functional of spherical 1PI amplitudes of a non-commutative 
scalar field theory. The derivation is done for $N\times N$ matrix field
theory but the result also applies to the Moyal case.

Let $F_\Lambda[J]$ be the functional of the $N\times N$ matrix source 
$J(p)$ that generates connected amplitudes. $F_\Lambda[J]$ is related
with the generating functional of connected and amputates amplitudes
$H_\Lambda[\varphi]$ via
\be
F_\Lambda[J] = H_\Lambda[\Delta_\Lambda\,J] + {1\over 2 N}\int\!\! d p \,
\Delta_\Lambda(p)\, {\rm Tr}\, J(p)\,J({\scriptstyle -}p)
\ee
and thus \cite{bgi} it satisfies the following finite $N$ RG equation
\bea
&&\Lambda\partial_\Lambda\,F_\Lambda = {1\over 2}\,\int\!\!d p\,\Ddot(p)\,
\Delta_\Lambda^{-2}(p)\,\Bigl[ N\,{\rm Tr}\,\Bigl(
{\delta F_\Lambda\over \delta J(p)}\,{\delta F_\Lambda\over \delta J({\scriptstyle -}p)}
\Bigr) \nonumber \\
&&\qquad\qquad\qquad + {1\over N}\,{\rm Tr}\,
{\delta^2 F_\Lambda\over \delta J(p)\,\delta  J({\scriptstyle -}p)} - \Delta_\Lambda(p)
\Bigr]
\label{Frg}
\eea
The generating functional $\Gamma_\Lambda^\prime[\varphi]$ of 1PI amplitudes
is the Legendre transform of $F_\Lambda[J]$:
\bea
&&\Gamma_\Lambda^\prime[\varphi] = 
\Gamma_\Lambda[\varphi] - {1\over 2 N}\,\int\!\! d p \,
\Delta^{-1}_\Lambda(p)\, {\rm Tr} \,\varphi(p)\,\varphi({\scriptstyle -}p)=\nonumber\\ 
&&\qquad\,\, =  F_\Lambda[J] - {1\over N}\,\int\!\!d p 
\,{\rm Tr}\, J(p)\,\varphi({\scriptstyle -}p)
\label{legendre1}
\eea
where
\be
\varphi_{\sss i j}(p)= N\, 
{\delta F_\Lambda \over \delta J_{\sss j i}({\scriptstyle -}p)}
\label{legendre2}
\ee
By taking the $\Lambda$ derivative of $\Gamma_\Lambda[\varphi]$ and using 
Eq.~(\ref{Frg}), one finds 
\bea
&&\!\!\!\!\!\!\!\!\Lambda\partial_\Lambda\,\Gamma_\Lambda = 
\Lambda\partial_\Lambda\,F_\Lambda - {1\over 2} \int\!\! d p \,\Ddot(p)\,
\Delta^{-2}_\Lambda(p)\, {\rm Tr} \,\varphi(p)\,\varphi({\scriptstyle -}p) =\nonumber\\
&&\qquad\,
= {1\over 2}\,\int\!\!d p\,\Ddot(p)\,
\Delta_\Lambda^{-2}(p)\,\Bigl[ {1\over N}\,{\rm Tr}\,
{\delta^2 F_\Lambda\over \delta J(p)\,\delta  J({\scriptstyle -}p)} - \Delta_\Lambda(p)
\Bigr]
\label{1PIfiniten1}
\eea
Let us introduce the matrices
\bea
&&\Bigl(\mathbb{F}^{(2)}_\Lambda\Bigr)_
{\sss (i_1 j_1;\,p_1),(i_2 j_2;\,p_2)}\equiv
{\delta^2 F_\Lambda\over \delta J_{\sss i_1 j_1}(p_1)\,\delta 
J_{\sss j_2 i_2}({\scriptstyle -}p_2)}\nonumber\\
&&\Bigl(\mathbb{G}^{\prime\,(2)}_\Lambda \Bigr)_
{\sss (i_1 j_1;\,p_1),(i_2 j_2;\,p_2)}\equiv
{\delta^2 \Gamma^\prime_\Lambda\over \delta 
\varphi_{\sss i_1 j_1}(p_1)\,\delta 
\varphi_{\sss j_2 i_2}({\scriptstyle -}p_2)}
\eea
whose row and column indices are given by the triples $(i j;\,p)$.  
Taking the $J$ derivative of 
Eq. (\ref{legendre2}) and the $\varphi$ derivative of (\ref{legendre1})
one obtains
\be
\mathbb{F}^{(2)}_\Lambda\,
\mathbb{G}^{\prime\,(2)}_\Lambda = \mathbb{I}
\label{inverses}
\ee
where, here and in the following, matrix multiplication involves both a sum
over double indices $(i j)$ and an integral over momentum $p$; furthermore 
$\mathbb{I}\equiv \delta_{i_1 i_2}\,\delta_{j_1 j_2}\, \delta(p_1-p_2)$.
The result (\ref{inverses}) together with (\ref{1PIfiniten1}) leads to the RG
evolution equation for the 1PI generating functional at finite $N$:
\bea
&&\!\!\!\!\!\!\!\!\!\Lambda\partial_\Lambda\,\Gamma_\Lambda = {1\over 2 N^2}\,
\mathbb{T}\mathrm{r}\,\Bigl[\dot{\mathbb{D}}_\Lambda\,
\mathbb{D}_\Lambda^{-1}\,
\Bigl(\mathbb{I}- N\,\mathbb{D}_\Lambda\,\mathbb{G}^{(2)}_\Lambda 
\Bigr)^{-1}\Bigr] - {1\over 2 N^2}\,\mathbb{T}\mathrm{r}\,\dot{\mathbb{D}}_\Lambda\,\mathbb{D}_\Lambda^{-1} =\nonumber\\
&&\qquad = {1\over 2}\sum_{n=1}^\infty N^{n-2}\,\mathbb{T}\mathrm{r}\,\Bigl[
\dot{\mathbb{D}}_\Lambda\,\mathbb{D}_\Lambda^{-1}\,
\Bigl(\mathbb{D}_\Lambda\,\mathbb{G}^{(2)}_\Lambda\Bigr)^n \Bigr]
\label{1PIfiniten2}
\eea
where 
\be
\bigl(\mathbb{D}_\Lambda\bigr)_{\sss (i_1 j_1;\,p_1),(i_2 j_2;\,p_2)}
\equiv  \Delta(p_1)\,\delta_{i_1 i_2}\,\delta_{j_1 j_2}\,\delta(p_1-p_2)
\ee
and $\mathbb{T}\mathrm{r}$ denotes the trace over the triple  $(i j;\,p)$.

The large $N$ limit of Eq.~(\ref{1PIfiniten2}) is defined by taking the
invariants
\be
\Omega_k (p_1\cdots
p_k)= {1\over N}\, {\rm Tr} \Bigl(\varphi (p_1)\cdots
\varphi (p_k)\Bigr)
\ee
fixed as $N\to\infty$. Hence one has to express derivatives with respect to 
$\varphi$ in terms of $\Omega_k$-derivatives:
\be
{\delta \Gamma_\Lambda \over \delta \varphi_{\sss i j}(p)} =
{1\over N}\,\sum_k k\,\int
\prod_{\a=1}^{k-1} d q_{\a}\,
\bigl(\varphi(q_1)\cdots\varphi(q_{k-1})\bigr)_{\sss j i}\,
{\delta \Gamma_\Lambda \over \delta 
\Omega_k(p,q_1,\ldots,q_{k-1})}
\ee
and
\bea
&&\!\!\!\!\!\!
{\delta^2 \Gamma_\Lambda\over \delta\varphi_{\sss i_1 j_1}(p_1)\,
\delta \varphi_{\sss j_2 i_2}({\scriptstyle -}p_2)} =
{1\over N}\sum_k\sum_{I=0}^{k-2} k\!
\int \!\prod_{\a=1}^{k-2} d q_{\a}\,\bigl(\varphi(q_1)\cdots\varphi(q_I)\bigr)_{\sss i_2 i_1}\times\\
&&\qquad
\times\bigl(\varphi(q_{I+1})\cdots\varphi(q_{k-2})\bigr)_{\sss j_1 j_2}\,
{\delta \Gamma_\Lambda \over \delta 
\Omega_k(p_1,q_1,\ldots,q_I,{\scriptstyle -}p_2,q_{I+1},\ldots,q_{k-2})}+\nonumber\\
&&\qquad + {1\over N^2}\, \sum_{k,k^\prime} k\,k^\prime\!\int\!
\prod_{\a=1}^{k-1} d q_{\a}\!\!\prod_{\b=1}^{k^\prime-1} d q^\prime_{\b}\,
\bigl(\varphi(q_1)\cdots\varphi(q_{k-1})\bigr)_{\sss j_1 i_1}\times\nonumber\\
&&\qquad\times \bigl(\varphi(q^\prime_1)\cdots\varphi(q^\prime_{k^\prime-1})
\bigr)_{\sss i_2 j_2}\,
{\delta^2 \Gamma_\Lambda \over \delta \Omega_k(p_1,q_1,\ldots,q_{k-1})\,
\delta \Omega_{k^\prime}({\scriptstyle -}p_2,q^\prime_1,\ldots,q^\prime_{k-1})}
\nonumber
\eea
The second addendum in the R.H.S. of the equation above if of sub-leading order
in $1/N$ and must be discarded in the large $N$ limit. Thus we find:
\bea
\label{nproduct}
&&\!\!\!\!\!\!\!\!
N^{n-2}\,\mathbb{T}\mathrm{r}\Bigr(\mathbb{G}^{(2)}_\Lambda\Bigr)^n =  
\int\! d{\rm P_0}\cdots d{\rm P_{n-1}}\,\prod_{i=1}^n
\Biggl[ \sum_{k_i}\sum_{I_i=0}^{k_i-2}\int\! \!
\prod_{\a_i=1}^{I_i} 
\!\!dp^{(i)}_{\a_i} \!\!\prod_{\b_i=1}^{k_i-2-I_i}\!\! 
dq^{(i)}_{\b_i}\times\\
&&\quad\times\,k_i\, {\delta
\Gamma_\Lambda\over\delta\Omega_{k_i}({\rm P_{i-1}}, C_i,
{\scriptstyle -}{\rm P_{i}}, C^\prime_i)}\Biggr]\,
\Omega_{\textstyle{\sss\sum_i I_i}}(C_n,\ldots, C_1)\,
 \Omega_{\textstyle{\sss\sum_i k_i-2-I_i}}(C^\prime_1,\ldots, C^\prime_n)\nonumber
\eea
where $C_i\equiv \{p^{(i)}_{\a_i}\}$, $C^\prime_i\equiv \{q^{(i)}_{\b_i}\}$.
Using the identity (\ref{nproduct}) in the flow equation (\ref{1PIfiniten2})
we end up with the large $N$ (or large $\theta$) RG equation for the 1PI
generating functional
\bea
&&\Lambda\partial_\Lambda \Gamma_\Lambda = {1\over2}\,\sum_{n=1}^\infty
\int\! d{\rm P_0}\cdots d{\rm P_{n-1}}\,\Ddot({\rm P_0})\,\Del({\rm P_1})\cdots
\Del({\rm P_{n-1}})\times\nonumber\\
&&\qquad\times
\prod_{i=1}^n\Biggl[ \sum_{k_i}\sum_{I_i=0}^{k_i-2}\int\! \!\prod_{\a_i=1}^{I_i} 
\!\!dp^{(i)}_{\a_i} \!\!\prod_{\b_i=1}^{k_i-2-I_i}\!\! dq^{(i)}_{\b_i}\, k_i\, {\delta
\Gamma_\Lambda\over\delta\Omega_{k_i}({\rm P_{i-1}}, C_i,
{\scriptstyle -}{\rm P_{i}}, C^\prime_i)}\Biggr]\times\nonumber\\
&&\qquad\times\, 
\Omega_{\textstyle{\sss\sum_i I_i}}(C_n,\ldots, C_1)\,
 \Omega_{\textstyle{\sss\sum_i k_i-2-I_i}}(C^\prime_1,\ldots, C^\prime_n)
\eea
which coincides with Eq.~(\ref{1PIlargenrg}) of Section 2.


\begin{thebibliography}{10}
\bibitem{seiberg} 
S.~Minwalla, M.~Van Raamsdonk and N.~Seiberg,
``Noncommutative perturbative dynamics,'' JHEP {\bf 0002} (2000) 020,
{\tt hep-th/9912072};
\\ M.~Van Raamsdonk and N.~Seiberg, ``Comments on
noncommutative perturbative dynamics,'' JHEP {\bf 0003} (2000) 035,
{\tt hep-th/0002186}.

\bibitem{arefeva} 
I.~Y.~Aref'eva, D.~M.~Belov and A.~S.~Koshelev,
``Two-loop diagrams in noncommutative phi**4(4) theory,''
Phys.\ Lett.\ B {\bf 476}, 431 (2000), {\tt hep-th/9912075}.

\bibitem{micu}
A.~Micu and M.~M.~Sheikh Jabbari,
``Noncommutative phi**4 theory at two loops,''
JHEP {\bf 0101}, 025 (2001), {\tt hep-th/0008057}.

\bibitem{bgi} 
C.~Becchi, S.~Giusto and C.~Imbimbo,
``The Wilson-Polchinski renormalization group equation in the planar  limit,''
Nucl.\ Phys.\ B {\bf 633}, 250 (2002),
{\tt hep-th/0202155}.

\bibitem{polchinski} J.~Polchinski, 
``Renormalization And Effective Lagrangians,''
Nucl.\ Phys.\ B {\bf 231} (1984) 269.

\bibitem{griguolo}
L.~Griguolo and M.~Pietroni,
``Wilsonian renormalization group and the non-commutative IR/UV  connection,''
JHEP {\bf 0105}, 032 (2001),
{\tt hep-th/0104217}.

\bibitem{ek}
T.~Eguchi and H.~Kawai,
``Reduction Of Dynamical Degrees Of Freedom In The Large N Gauge Theory,''
Phys.\ Rev.\ Lett.\  {\bf 48}, 1063 (1982).

\bibitem{theisen}
A.~Armoni, E.~Lopez and S.~Theisen,
``Nonplanar anomalies in noncommutative theories and the Green-Schwarz  
mechanism,'' JHEP {\bf 0206}, 050 (2002), {\tt hep-th/0203165}.


\end{thebibliography}
\end{document}